\documentclass[12pt]{article}

\usepackage{times}
\usepackage{graphicx}
\usepackage{hyperref}
\usepackage{authblk}
\usepackage{todonotes, xcolor, soul}
\usepackage{lineno}
\usepackage[super, sort&compress]{natbib}
\usepackage{bm} 

\newenvironment{sciabstract}{%
\begin{quote} \bf}
{\end{quote}}

\title{Observation of competing, correlated ground states in the flat band of rhombohedral graphite}

\author[2, 6, 7, 8 $\dagger$]{Imre Hagym\'{a}si}
\author[1, $\dagger$]{Mohammad Syahid Mohd Isa}
\author[1, 5]{Zolt\'{a}n Tajkov}
\author[1]{Kriszti\'{a}n M\'{a}rity}
\author[3, 4]{Oroszl\'{a}ny L\'{a}szl\'{o}}
\author[5]{J\'{a}nos Koltai}
\author[3]{Assem Alassaf}
\author[1]{P\'{e}ter Kun}
\author[1]{Konr\'{a}d Kandrai}
\author[1]{Andr\'{a}s P\'{a}link\'{a}s}
\author[1]{P\'{e}ter Vancs\'{o}}
\author[1]{Levente Tapaszt\'{o}}
\author[1, *]{P\'{e}ter Nemes-Incze}

\affil[1]{Centre for Energy Research, Institute of Technical Physics and Materials Science, 1121 Budapest, Hungary}
\affil[2]{Helmholtz-Zentrum Berlin f\"{u}r Materialien und Energie, 14109 Berlin, Germany}
\affil[3]{Department of Physics of Complex Systems, ELTE Eötvös Loránd University, 1117 Budapest, Hungary}
\affil[4]{Budapest University of Technology and Economics, 1111 Budapest, Hungary}
\affil[5]{ELTE E\"{o}tv\"{o}s Lor\'{a}nd University, Department of Biological Physics, 1117 Budapest, Hungary}
\affil[6]{Wigner Research Centre for Physics, 1121 Budapest, Hungary}
\affil[7]{Dahlem Center for Complex Quantum Systems and Institut f\"{u}r Theoretische Physik, Freie Universit\"{a}t Berlin, 14195 Berlin, Germany}
\affil[8]{Max Planck Institute for the Physics of Complex Systems, Dresden, Germany}
\affil[$\dagger$]{\small \emph{These authors contributed equally}}
\affil[*]{\small \emph{corresponding author, email: nemes.incze.peter@ek-cer.hu}}

\date{}

\begin{document} 


\baselineskip24pt


\maketitle 



\newpage


\begin{sciabstract}
In crystalline solids the interactions of charge and spin can result in a variety of emergent quantum ground states, especially in partially filled, topological flat bands such as Landau levels or in 'magic-angle' bilayer graphene.
Much less explored is rhombohedral graphite (RG), perhaps the simplest and structurally most perfect condensed matter system to host a flat band protected by symmetry.
By scanning tunneling microscopy we map the flat band charge density of 8, 10 and 17 layers and identify a domain structure emerging from a competition between a sublattice antiferromagnetic insulator and a gapless correlated paramagnet.
Our density-matrix renormalization group calculations explain the observed features and demonstrate that the correlations are fundamentally different from graphene based magnetism identified until now, forming the ground state of a quantum magnet.
Our work establishes RG as a new platform to study many-body interactions beyond the mean-field approach, where quantum fluctuations and entanglement dominate.
\end{sciabstract}


\section*{Introduction}

An important factor separating conventional and quantum magnets is that in the latter the mean-field description fails due to strong correlations and quantum fluctuations.
Examples include antiferromagnetic spin chains~\cite{Haldane1983-sb} and ladders~\cite{Dagotto1996-yv}, where strong interactions result in exotic properties such as fractionalization and topological order~\cite{Schollwock2004-ld}.
Generally, condensed matter systems realizing quantum magnets have a complex structure, requiring the spins provided by the \emph{d}-band of transition metals~\cite{Vasiliev2018-hg}.
On the other hand, graphene based materials comprising only carbon atoms are simple and host a remarkable multitude of magnetic states, provided that a band with a divergent charge density is partially filled.
This divergence can arise through a sublattice imbalance~\cite{Magda2014-ew,Just2014-go,Mishra2021-mr,Lemonik2012-rp,Lang2012-qi}, a Landau level~\cite{Young2012-tz,Young2013-wh}, or in a moir\'{e} superlattice~\cite{Sharpe2019-uu,Chen2020-vz}.
Perhaps the simplest example of this is at the edges of zigzag nanoribbons, where at charge neutrality~\cite{Vancso2017-qx}, the spins on opposite sublattices interact antiferromagnetically~\cite{Magda2014-ew}.
Further examples are in the layer antiferromagnet state of bilayer graphene~\cite{Lang2012-qi,Velasco2012-ja,Geisenhof2021-ag} and in the sublattice-N\'{e}el state in trilayer~\cite{Lee2014-zc,Yankowitz2014-wm,Hattendorf2013-fh} and tetralayer RG~\cite{Myhro2018-me,Lee2019-cu,Kerelsky2021-os}, all explained within the mean-field approximation.
This N\'{e}el state has the spins on opposing sublattices in an antiparallel configuration between the top and bottom surface and in a ferrimagnetic orientation within the same graphene layer.
Mean-field Hubbard models predict it to be the gapped ground state at charge neutrality~\cite{Xu2012-bw,Muten2021-cb}, measured in three and four layer RG by transport~\cite{Lee2014-zc} and STM~\cite{Kerelsky2021-os}.
However, for thicker RG interactions are expected to increase~\cite{Pamuk2017-mj}, holding the potential for driving a transition to a quantum magnet, providing the first realization of graphene based magnetism beyond the mean-field model.

Here we show that for thick RG, the gapped N\'{e}el state is not the ground state, because fluctuation terms and local correlations can not be neglected.
Instead, RG hosts domains of alternating gapped and gapless surface charge density, associated with a degenerate ground state.
This degeneracy is a signature of quantum magnetism~\cite{Schollwock2004-ld} that can not be described at the mean-field level.

\begin{figure}[ht!]
  \includegraphics[width = 1 \textwidth]{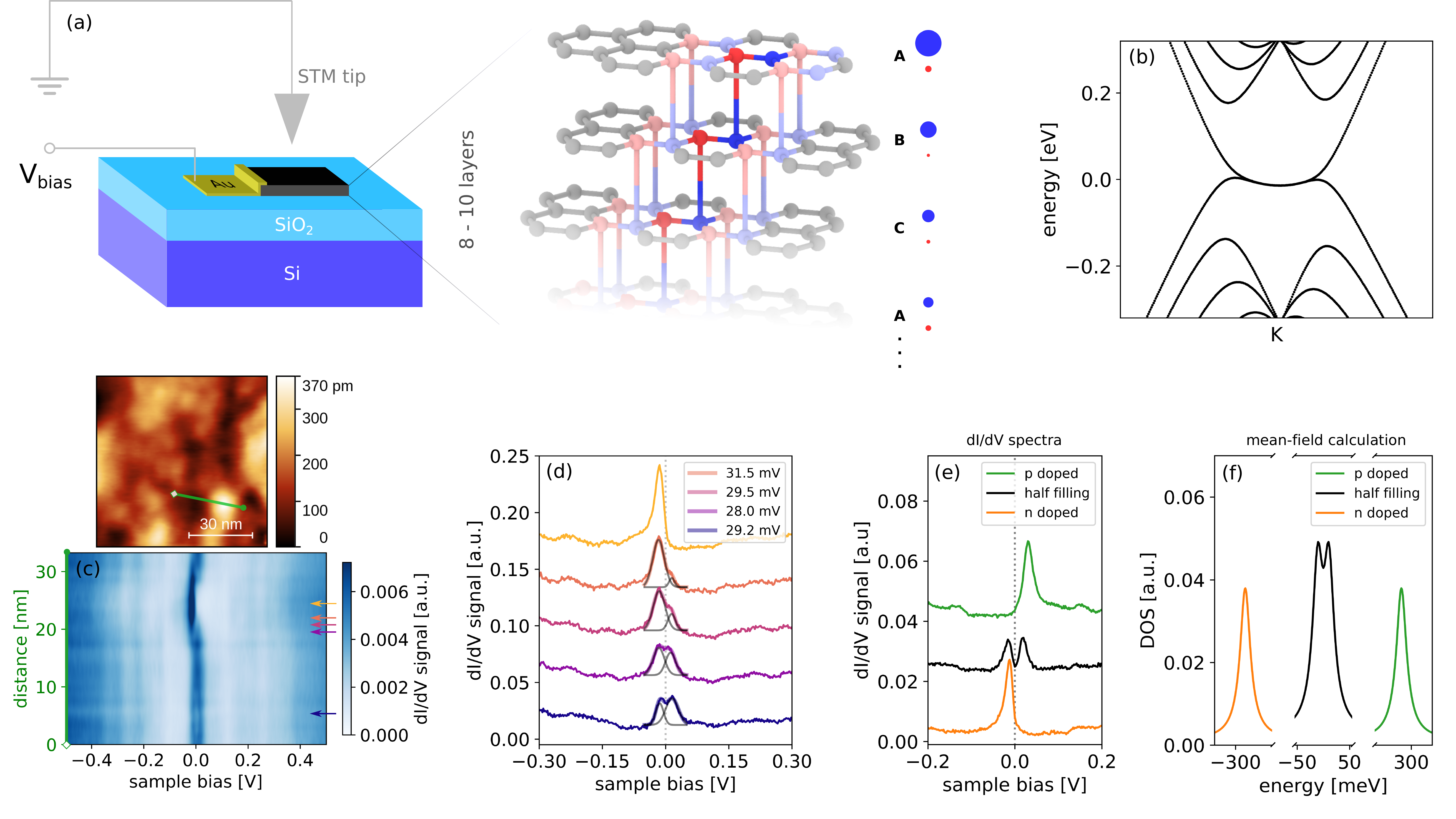}
  \caption{\textbf{Correlated insulating state in RG.}
    \textbf{(a)} Schematic of the STM measurement setup, with the RG supported on a SiO$_2$ substrate.
    Blue and red atoms/bonds, show the staggered, SSH-like~\cite{Su1979-ok} chain in RG~\cite{Min2008-zp}.
    Nearest neighbor chains, shown in lighter blue and red colors form a ladder, through inter-chain hoppings.
    The size of the blue and red circles next to the graphene layers is proportional to the LDOS on the respective sublattices.
    A, B and C mark the characteristic stacking pattern of RG.
    \textbf{(b)} Calculated ab-initio band structure of RG around the K point, for 8 layers.
    \textbf{(c)} Density plot of 64 $dI/dV$ spectra measured along the green line shown in the STM topography image above (8 layer sample).
    The surface state shifts in energy from being at the Fermi level to being almost completely filled and back to half filling.
    \textbf{(d)} Selected spectra, positions shown by arrows of the same color as in (c).
    The splitting in the surface state is always at the Fermi level, shown by the dotted line.
    Black Gaussians are fits to the surface state $dI/dV$ peak, with a splitting shown in the inset.
    \textbf{(e)} $dI/dV$ spectra measured in various positions on the sample surface, with the surface state below (orange), above (green) and at the Fermi level (black).
    The surface state shows a splitting of 32 meV at charge neutrality (black).
    The FWHM of the completely filled or empty surface state is 26 meV.
    \textbf{(f)} Surface total DOS, calculated within the mean-field approach for various doping levels ($U$ = 6 eV).
    Colors with the same meaning as in (e).
  }
\label{fig:intro}
\end{figure}

RG is distinct from any other graphene system in that it features a staggered intra and inter-layer hopping pattern~\cite{Min2008-zp}, similar to a 1D Su-Schrieffer-Heeger model~\cite{Su1979-ok}.
It can be considered as built up from parallel SSH chains  (Fig.~\ref{fig:intro}a) leading to a sublattice imbalance on the top and bottom surfaces.
A flat band is localized on the unpaired surface sublattice (blue in Fig.~\ref{fig:intro}a), which decays into the bulk~\cite{Xiao2011-ax,Slizovskiy2019-un}.
The interacting 1D SSH chain~\cite{Nawa2019-ko,Le2020-zz} and "ladder materials"~\cite{Dagotto1996-yv} with the chains linked up parallel to each other are elementary models for quantum magnets.
This, together with the recent evidence for strong correlations in thick samples~\cite{Shi2020-bv} make RG a prime candidate for quantum magnetism.

Scanning tunneling microscopy (STM) is a powerful tool to investigate correlated electron systems, for example uncovering the inhomogeneities in the order parameter of high $T_{\mathrm{C}}$ superconductors~\cite{Dagotto2005-bn}, or nematicity in "magic-angle" bilayer graphene~\cite{Kerelsky2019-gg,Jiang2019-ob,Choi2019-ym,Xie2019-kd}.
We present STM measurements at a temperature of 9.6 K, on the surface of SiO$_2$ supported RG (Fig.~\ref{fig:intro}a).
At half filling of the flat band (charge neutrality) two distinct phases of the surface state charge density appear.
One having a splitting of up to 40 meV centered on the Fermi level, the other state being gapless.
The two phases form a domain structure, characteristic of many-body competing ground states~\cite{Dagotto2005-bn}.
Using density matrix renormalization group (DMRG) calculations, we identify that the surface hosts a net spin of $\mathcal{S} = 1$.
This magnetic moment gives rise to a ground state of RG, that has the degenerate spin projections on the surface of: $s_z = \pm1$ and $s_z = 0$.
The $s_z = \pm1$ surface state corresponds to the sublattice-N\'{e}el insulator, identified earlier using mean-field approaches: ab-initio~\cite{Pamuk2017-mj}, Hubbard~\cite{Lee2014-zc}, Hartree-Fock~\cite{Jung2013-ks} and continuum model~\cite{Zhang2011-wu} calculations.
The $s_z = 0$ surface state is a correlated paramagnet, that is not present at the mean-field level~\cite{Jung2013-ks,Lee2014-zc}.
The degeneracy between the gapless paramagnetic state and the antiferromagnetic insulator is an intrinsic property of RG and explains the domain structure of the surface state splitting observed experimentally.

\section*{Results}
\subsubsection*{Gapped and gapless domains at charge neutrality}

We have used Raman spectroscopy to identify thick RG samples on a SiO$_2$ support (for details see SI1.1) and performed STM measurements at multiple locations on an 8, 10, 14 and 17 layer sample.
The data presented in the main text is representative of the behavior seen in all samples.
A topographic image of an $80\times80$ nm area of the 8 layer thick sample is shown in Figure~\ref{fig:intro}c, while the large density of states of the surface flat band shows up as a pronounced peak in the $dI/dV$ spectra (Fig.~\ref{fig:intro}c,d,e).
Alongside this peak we can observe step-like features in the $dI/dV$ signal starting below -150 mV and above +150 mV.
These stem from dispersive, bulk bands visible in Fig.~\ref{fig:intro}b and their position in energy is a fingerprint of the number of layers in the RG sequence (see SI3).
By tracking these peaks in the areas we investigate, we can ensure that the stacking sequence of RG is the same all over the measured surface (Fig. S9d).
The RG sample is only perturbed by the local charge and mechanical deformation inhomogeneity of the SiO$_2$ support.
It is well known that this modulates the doping in the sample placed on top~\cite{Zhang2009-gn} and
we use this inhomogeneity to our advantage.
By simply moving the STM tip over the sample we are able to investigate the surface state at various electron fillings.

Figure~\ref{fig:intro}c, shows spectra across a 35 nm long line on the surface.
The surface state suffers a local shift across the Fermi level (zero bias), with selected spectra shown in Fig.~\ref{fig:intro}d.
We observe a $\sim$30 meV splitting, centered on zero bias as soon as the surface state becomes partially filled with electrons, which is a strong indication of its many-body origin.
In Fig.~\ref{fig:intro}e, we highlight the three representative cases, with the surface state at charge neutrality (black), filled (orange) and empty (green).
In the latter two situations, the full width at half maximum (FWHM, $w$) is 26 meV, the narrowest peak width we measure.
In this case it is usually assumed that the effect of interactions is negligible.
However, the filled and empty peaks show a shoulder on the right and left sides respectively.
This asymmetry is characteristic of the doped spectra, similarly to the doped flat band peak of "magic angle" bilayers, indicating that e-e interactions are present even near complete filling of the band~\cite{Xie2019-kd,Kerelsky2019-gg}.
At half filling, the $dI/dV$ peak shows a splitting of 32 meV.
To reveal the origin of the observed peak splitting we have performed calculations at the mean-field level, evidencing the opening of a gap due to antiferromagnetic exchange~\cite{Lee2014-zc} (see Fig.~\ref{fig:intro}f and SI2 for details).
To rule out single particle effects as the origin of the surface state splitting, we have modeled the effect of a perpendicular electric field and quantum confinement due to the electrostatic disorder potential of the sample.
Using a density functional theory and a tight binding model, we found that these effects are not consistent with our observations either qualitatively or quantitatively.
Splittings induced by quantum confinement are not pinned to the Fermi level, therefore are inconsistent with our observation.
Furthermore, the perpendicular electric field needed to open a gap of 30 meV is orders of magnitude larger then what is present in our experiment.
For a detailed discussion see section SI8 of the supplement.

\begin{figure}[h]
  \begin{center}
  \includegraphics[width = 0.9 \textwidth]{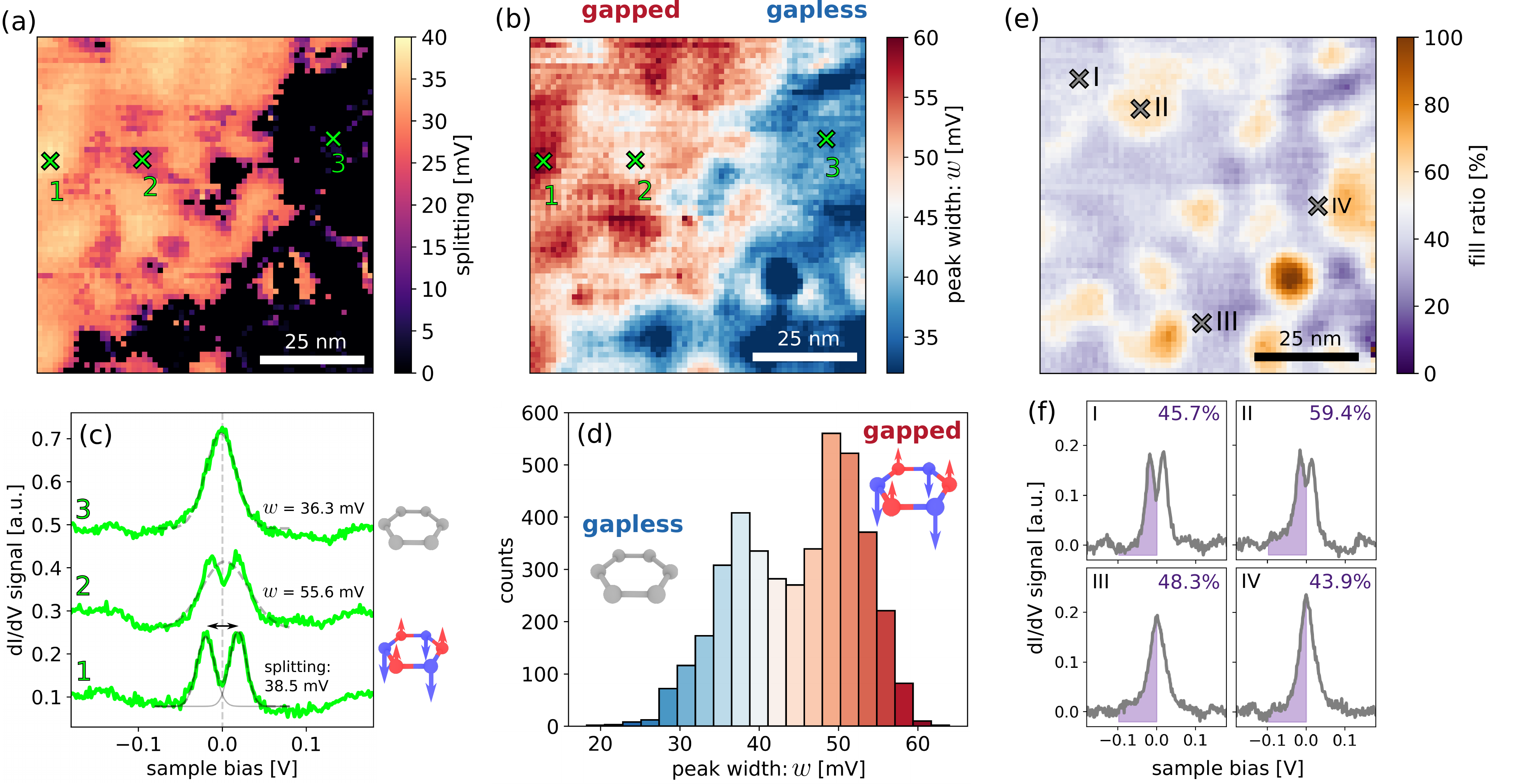}
  \caption{\textbf{Gapped and gapless domains.}
    \textbf{(a)} Map of the surface state splitting over an 80$\times$80 nm area of the sample, measured using a grid of 64$\times$64 $dI/dV$ spectra.
    STM topographic image of the region can be seen in Fig.~\ref{fig:intro}c.
    The splitting was determined by fitting two Gaussians to the peak.
    Examples of fits can be seen in (c) and Fig.~\ref{fig:intro}d.
    \textbf{(b)} FWHM ($w$) of the peak from the same $dI/dV$ map, obtained by fitting a single Gaussian (for fitting details see SI4).
    \textbf{(c)} Examples of individual $dI/dV$ spectra measured in locations shown by numbers on (a) and (b).
    Double and single Gaussian fits shown by gray dashed curves, the splitting and $w$ values shown next to the spectra.
    \textbf{(d)} Histogram of peak widths ($w$) in the data of (b).
    \textbf{(e)} Map of the local fill ratio as defined in the main text for the area shown in (a).
    \textbf{(f)} Selected spectra measured at the positions marked by crosses and roman numerals in (e).
    Purple shading shows the area under the peak used to determine the fill ratios shown in the inset in \%.
    Tip stabilization parameters: 100 pA, 500 mV.
  }
  \label{fig:splitting}
  \end{center}
\end{figure}

Next, we have quantified the local variation of the splitting over large areas of the sample surface.
By fitting two Gaussians to the surface state peak, we can extract the local splitting as the voltage difference between the peaks of the two Gaussians.
The fitting was performed in the $\pm$80 mV interval around zero bias.
Examples of such fits can be seen in Fig.~\ref{fig:intro}d and~\ref{fig:splitting}c.
In Figure~\ref{fig:splitting}a we show a representative map of the local splitting in an 80$\times$80 nm area.
Details of the fitting procedure and more splitting maps are shown in sections SI4 and SI5.2 of the supplement.
One striking feature of the map is that the sample contains regions with no observable splitting, even though the flat band is half filled in most of the area, as illustrated by spectrum \#3, \#II and \#IV in Fig.~\ref{fig:splitting}.
The local charge density can be approximated by the spectral weight of the surface state below and above the Fermi level, similarly to the procedure used by Jiang et al. in the case of "magic angle" bilayer graphene~\cite{Jiang2019-ob}.
After subtracting the background from the $dI/dV$ spectra, we measure the area of the peak under the Fermi level ($A_{under}$) up to the peak edge (-100 mV).
Examples of this can be seen in Fig.~\ref{fig:splitting}f.
We define the \emph{fill ratio} as $A_{under} / A_{total}$, where $A_{total}$ is the total area under the peak.
We consider half filling, or charge neutrality to be at 50\%\ fill ratio.
The local charge density measured this way is also correlated with the shift in bias voltage of the bulk bands (for more details see SI5.1).

A map of the local fill ratio (Fig.~\ref{fig:splitting}e) shows that in most areas the peak is between 30\%\ to 60\%\ filled.
There are significant areas of near half filling in both domains with large splitting and no splitting at all.
This is surprising at first, since the mean-field ground state in tri- and tetra-layers is gapped~\cite{Lee2014-zc,Myhro2018-me,Kerelsky2021-os}.
The presence of a gapless surface density indicates the failure of the mean-field approximation to fully capture the behavior of 8 layer RG.

The domain structure is also apparent if we fit a single Gaussian to the spectra and plot the FWHM $w$, which shows a bimodal distribution (Fig.~\ref{fig:splitting}d).
Furthermore, the peak width in the areas without splitting is at least 10 mV larger than in the fully doped case, suggesting that many-body effects do play a role in this case.
We call the areas without splitting: gapless, up to the energy resolution of our measurement, meaning that we can not rule out the presence of a gap smaller than 4.5 meV.
In the measurements, slight deviation from charge neutrality, but with the Fermi levels still near half filling, does not alter the gapped or gapless nature of the peak, as can be seen in Fig.~\ref{fig:intro}d and~\ref{fig:splitting}.

\subsubsection*{Degenerate ground state in RG and the breakdown of the mean-field approximation}

\begin{figure}
  \begin{center}
  \includegraphics[width = 1 \textwidth]{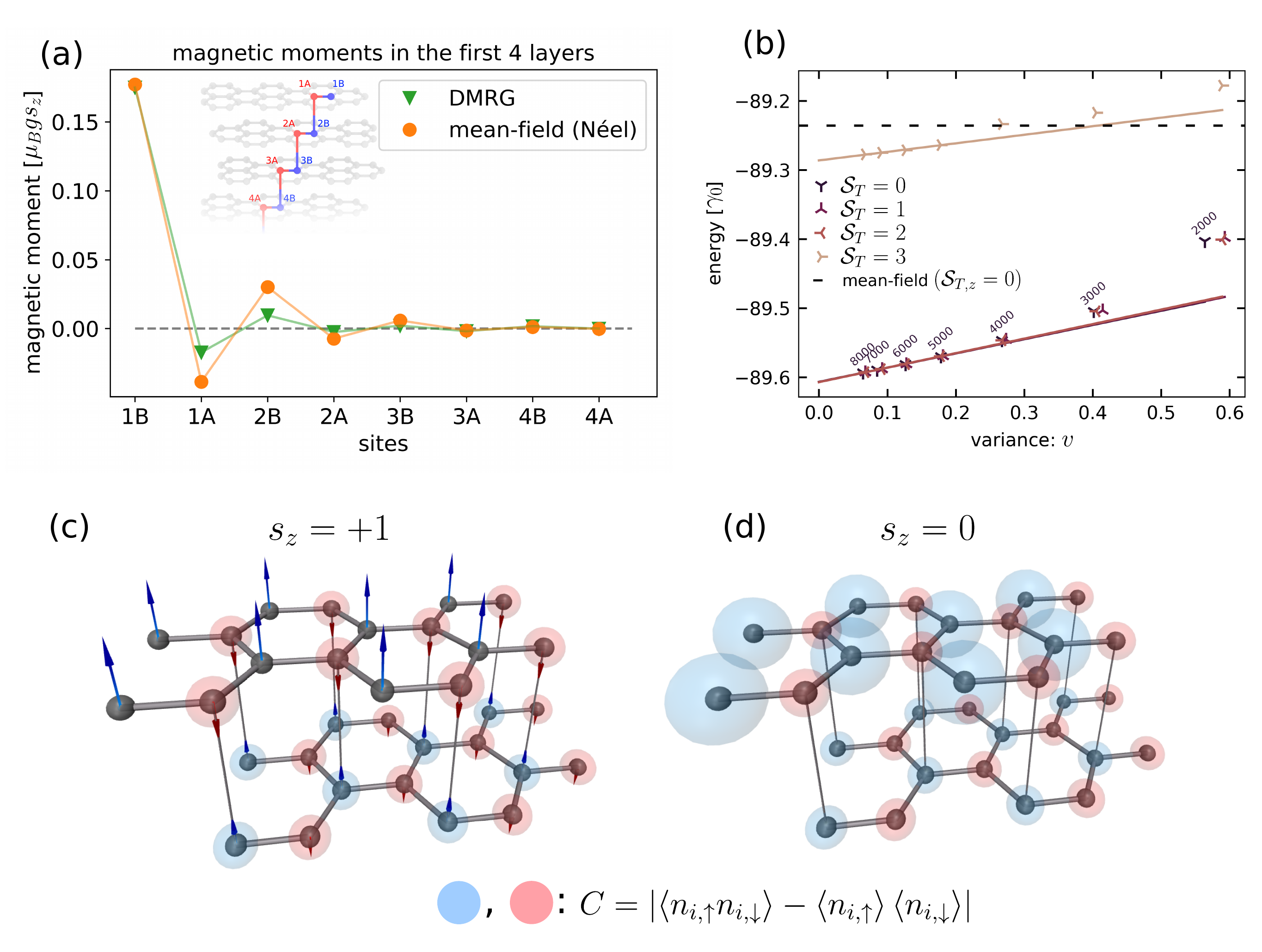}
  \caption{\textbf{DMRG and mean-field Hubbard model calculations.}
    \textbf{(a)} Calculated local magnetic momenta for a 6 layer RG.
    The magnetic moments on the atomic sites decay exponentially, essentially to zero after the second graphene layer ($U = 0.5 \times \gamma_0$).
    Inset shows one staggered chain of RG in red and blue atoms, along which the plot is performed.
    \textbf{(b)} DMRG calculations of the ground states of a 6 layer thick cell as a function of the variance.
    Energy in units of the nearest neighbor hopping term: $\gamma_0$.
    The exact ground state is reached at zero variance.
    The states with various total spin of whole cell (both top and bottom graphene layers): $\mathcal{S}_T$ are degenerate, with a ground state that is 5.16 meV/atom smaller than the mean-field case (dashed line).
    The first excited spin state ($\mathcal{S}_T$ = 3) is 0.33$\times \gamma_0$.
    \textbf{(c, d)} Colored arrows showing the distribution of magnetic moments on the top two graphene layers with a surface spin of $s_z$, from the DMRG calculation.
    The magnetic moments $\mu$ are scaled by $2|\mu|^{1/4}$, for better visibility.
    The radius of the opaque spheres is proportional to the local electronic correlation $C$ values, as defined by the relation shown.
    $\left < n_{i, \uparrow} \right >$ and $\left < n_{i, \downarrow} \right >$ are the expectation values of the spin density on the atomic site $i$.
    Blue and red colors distinguish the two graphene sublattices.
  }
  \label{fig:theory}
  \end{center}
\end{figure}

The dual nature of the flat band charge density is remarkable, since at the mean-field level, gapless ground states are not expected to be stable \cite{Jung2013-ks}.
The DMRG approach goes beyond the mean-field approximation and allows us to compute the complete many-body wave function, taking quantum fluctuations fully into account that are crucial in low-dimensional systems.
Fig.~\ref{fig:theory}a shows the magnetic moments on the top four graphene layers along a staggered chain of RG, calculated using DMRG.
The local moments show an antiferromagnetic ordering on the graphene sublattices, which decay into the bulk to near zero by the third graphene layer.
It can be seen that the mean-field calculations also reproduce this sublattice-N\'{e}el magnetic ordering, where the total spin for one surface, $s_z$ within the RG unit cell is 1.
Our mean-field calculations reproduce this N\'{e}el ordering, which forms an insulating state~\cite{Xu2012-bw,Muten2021-cb}, with a splitting shown in Fig.~\ref{fig:intro}f.
The splitting observed in the $dI/dV$ maps is a signature of this insulating state.
One expects an insulator, because the magnetic ordering breaks the inversion and time reversal symmetries protecting the gapless state at charge neutrality~\cite{Manes2007-md}.

Beyond the agreement on the magnetic ordering, we find that the mean-field approach overestimates the ground state energy of RG by 5.16 meV/atom compared to the DMRG result, highlighting the importance of correlation effects in the system (for more details see SI2).
In DMRG calculations the ground state of the RG slab features spins localized to the bottom and top layers, with a total spin of $\mathcal{S}_T$ and a spin projection of $\mathcal{S}_{T,z}$.
In Fig.~\ref{fig:theory}b we plot the ground state energies, as a function of the two-site approximation of the full variance $v$, defined by $v = \left\langle H^2\right\rangle-\langle H\rangle^2$, where $H$ is the Hamiltonian and $\langle \dots \rangle$ denotes the expectation value with respect to the variational state.
Three different configurations of the total spin, $\mathcal{S}_T = 0, 1, 2$ converge
to the exact ground state at $v = 0$, all of them being degenerate.
Since in STM we are probing only one surface, these total spin states correspond to three possible spin projections on the top surface: $s_z = 0, \pm1$.
Among these, the $s_z = \pm1$ (N\'{e}el) also features in the mean-field result, with the $+1$ state shown in Fig.~\ref{fig:theory}c and the $-1$ state corresponding to the case, where the spins are inverted.
These two states are indistinguishable if the STM tunneling current is not spin polarized.
However, the $s_z = 0$ instance has no analogue in the mean-field approximation, since the mean-field decoupling excludes the possibility to host a correlated state and only magnetic solutions can result in a lower energy.
This correlation effect is shown in Fig.~\ref{fig:theory}c and d, where the radius of the colored spheres is proportional to the calculated charge correlation values ($C$).
Instead of the magnetic moments, the {$s_z = 0$} solution shows an enhanced local charge correlation on the unpaired sublattice, forming a correlated paramagnet on the surface.
In this state the charge fluctuations are more and more suppressed with increasing $U$ (the electrons prefer not to hop to already occupied sites) and the total energy is decreased directly by the smaller number of doubly occupied sites instead of long-range magnetic ordering.
This is a truly many-body effect, since the mean-field approach is only able to decrease the total energy via magnetic ordering and not by decreasing the doubly occupied sites.
This state is gapless for the charge degree of freedom, since there is no large scale breaking of symmetries~\cite{Manes2007-md}, as opposed to the $s_z = \pm1$ case.
The two degenerate states form a natural candidate for the domain structure observed in STM, with a gapped antiferromagnetic insulator and a gapless, correlated paramagnet.
Globally over the whole sample surface none of the two states can dominate, but locally due to inhomogeneities the system can favor either the $s_z = \pm1$ or the $s_z = 0$ state, similarly to competing states in strongly correlated materials~\cite{Dagotto2005-bn}.

Finally, we have calculated the energy of the $\mathcal{S}_T = 3$ state and find that it is $0.33 \times \gamma_0$ higher then the ground state for a 6 layer thick cell (Fig.~\ref{fig:theory}b), with $\gamma_0$ being the nearest neighbor hopping of graphene.
This gap separates, the first (bulk) excited spin state of RG, from the degenerate ground state.
The large size ($\sim$1 eV) of the spin gap is responsible for the rapid decay of the magnetic moments ($s_z$) at the top and bottom graphene layers.

\begin{figure}
  \includegraphics[width = 1 \textwidth]{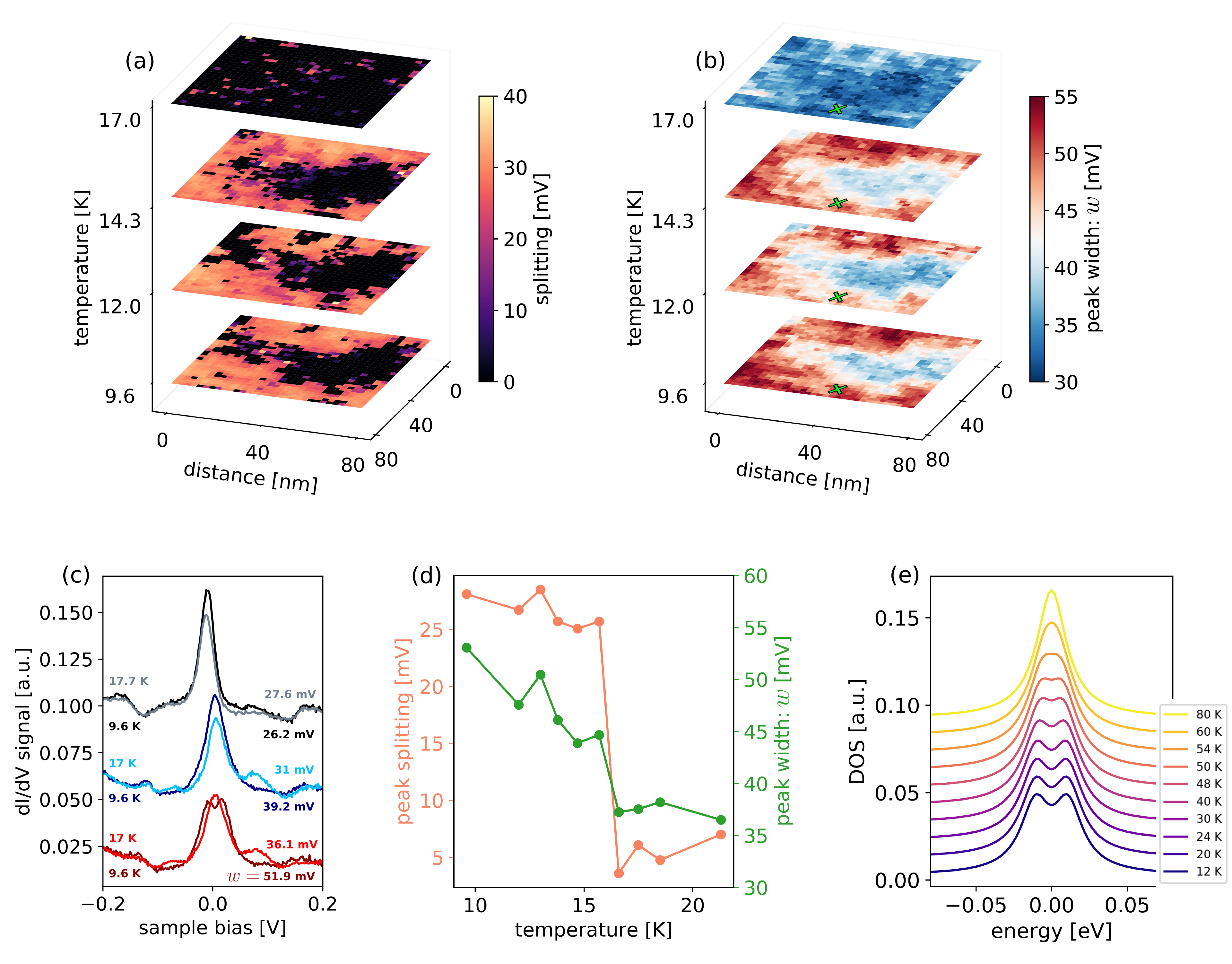}
  \caption{\textbf{Temperature dependence.}
    \textbf{(a)} Map of the surface state splitting in an $80 \times 80$ nm area on the 8 layer sample.
    The same area is measured at four different temperatures.
    \textbf{(b)} Map of $w$ in the same area and temperatures as in (a).
    Above 16.6 K, the splitting vanishes and the peak width decreases sharply.
    \textbf{(c)} Selected spectra in a gapped (bottom), gapless (middle) and completely filled (top) area below and above the critical temperature.
    The width: $w$ of the peaks shown in mV next to the spectra.
    Spectra at different temperatures are measured in the same location.
    \textbf{(d)} Peak splitting (orange) and $w$ of the surface state (green) as a function of temperature, measured in the position marked by the green crosses in (b).
    Connecting lines are guide to the eye.
    \textbf{(e)} Temperature dependent mean-field calculation of the splitting at charge neutrality (U = 6 eV).
  }
\label{fig:tc}
\end{figure}

\subsubsection*{Critical temperature}

Having established the presence of correlated insulating and gapless phases, we have checked their temperature dependence.
In Fig.~\ref{fig:tc}a and b, we show another example of the domain structure of the surface state splitting.
By measuring the same $80 \times 80$ nm area of the sample at increasing temperature, we observe that the splitting and $w$ suddenly decreases at a temperature of 17 K all across the surface.
Example spectra are shown in Fig.~\ref{fig:tc}c in a gapped (red) and a gapless (blue) region, measured at different temperatures in the same surface position.
Raising the temperature from 9.6 K to 17 K, the surface state becomes a single peak with decreasing $w$.
This is in contrast to a highly doped area (gray spectra), where $w$ increases by 1.4 mV, corresponding to the extra broadening due to the increased temperature (see also Fig. S13).

This suggests the presence of a phase transition at charge neutrality.
We identify as the critical temperature, the value where the splitting collapses and the width of the peak stops decreasing and plateaus.
Values measured on different areas of the sample show a variation.
The value of $T_{\mathrm{C}}$ for the area shown in Fig. \ref{fig:tc} is 16.6 K, with the highest $T_{\mathrm{C}}$ measured to be 22 K (see SI6).
Above $T_{\mathrm{C}}$, the $dI/dV$ peak width is still roughly 10 meV wider than the $w$ expected from single particle band structure (26 meV), suggesting that some many-body effects might still play a role at these temperatures.
Our temperature dependent mean-field calculation reproduces the narrowing of the peak by increasing temperature, but as expected, it overestimates $T_{\mathrm{C}}$ (Fig.~\ref{fig:tc}e).

Having determined the splitting and bandwidth in our sample, it is instructive to compare it to STM measurements of a 4 layer RG sample and to the correlated insulator states observed in "magic angle" twisted bilayer graphene.
In the former case, measurements by Kerelsky et al.~\cite{Kerelsky2021-os} show a FWHM of the surface state peak between 3 and 5 meV, while the splitting is 9.5 meV at half filling.
In our measurement of the 8 layer RG the FWHM is 25 meV, with a splitting in the 30 to 40 meV range.
Thus, the rearrangement of the spectral weight (splitting) due to interactions exceeds the bandwidth (FWHM) of the peak, which is a clear sign of strong correlations.
By this measure, 8 layer RG is also comparable to "magic angle" bilayers, where the splitting between the two flat band $dI/dV$ peaks increases by 15 to 20 meV near charge neutrality, also exceeding the bandwidth of around 10 meV~\cite{Kerelsky2019-gg,Xie2019-kd}.

Measurements of the surface state charge density in a 14 layer RG sample were also carried out by Henck et al.~\cite{Henck2018-js}, using nano-ARPES measurements.
Below the Fermi level they measure a FWHM of the surface state of 50 meV at a temperature of 90 K.
In their discussion the authors assume that the sample is charge neutral, with a surface state splitting of 40 meV surviving up to the measurement temperature.
The value of the splitting is comparable to our measurements, but their assumption that the critical temperature is higher than 90 K is not consistent with our STM data.

\begin{figure}
  \begin{center}
  \includegraphics[width = 0.6 \textwidth]{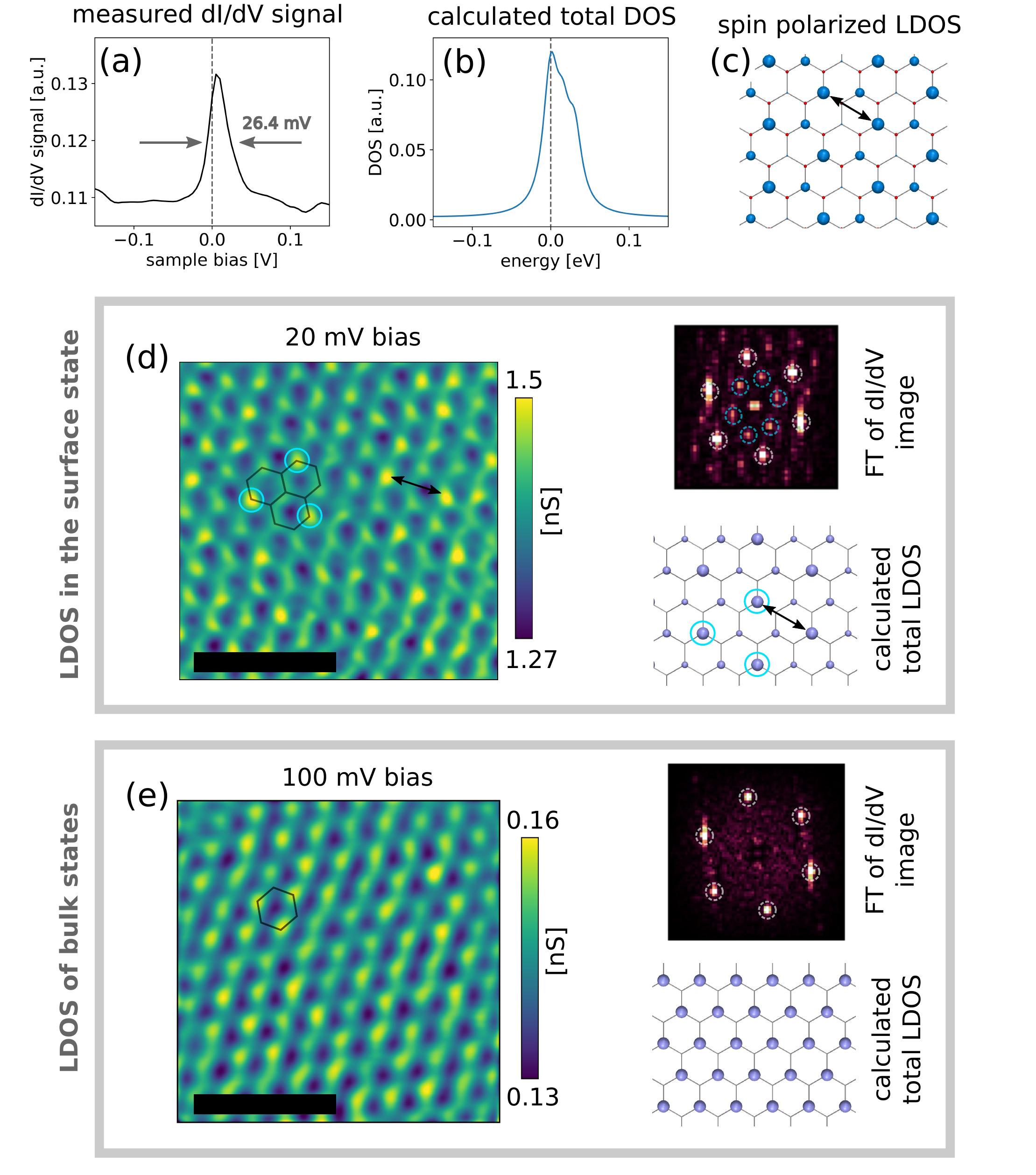}
  \caption{\textbf{Possible ferromagnetism and spontaneous breaking of translational symmetry.}
    \textbf{(a)} $dI/dV$ spectrum measured in an area with slight $p$ doping.
    FWHM, $w$ of the surface state peak is 26.4 mV.
    \textbf{(b)} DOS for $p$ doping of 2.16$\times 10^{12}$ cm$^{-2}$, similar to the experiment, calculated by the mean-field model.
    \textbf{(c)} Mean-field calculation of the spin resolved LDOS of the top layer of $p$ doped RG.
    Size of circles is proportional to the LDOS of the top graphene layer in the flat band ($U = 4.2$ eV).
    The LDOS shows essentially a ferromagnet, blue: up spin, red: down spin.
    Black arrow marks the $\sqrt{3} \times$ enlarged unit cell size.
    This magnetization pattern is also reproduced in DMRG.
    \textbf{(d)} STM of the $dI/dV$ signal inside the surface state.
    Black arrows and light blue circles show the enlarged unit cell size.
    Stabilization parameters: 30 pA, 20 mV.
    \textbf{(e)} $dI/dV$ image far from the surface state, measured in the same area as (d).
    Stabilization parameters: 70 pA, 100 mV.
    Insets in (d), (e): Fourier transform of the $dI/dV$ image.
    White, dashed circles show the atomic periodicity corresponding to the 2.46 \AA\ unit cell of graphene.
    Light blue, dashed circles show the unit cell of 2.46$\times \sqrt{3}$ \AA.
    Size of blue spheres is proportional to the calculated total LDOS in the flat band: (d) and 240 meV above the flat band: (e).
    Scale bars on the $dI/dV$ images are 1 nm in length.
    Stabilization parameters in (d) and (e) are chosen, such that the tip - sample distance stays the same.
    Measurement temperature: 9.6 K, bias modulation: 20 mV.
  }
  \label{fig:ferro}
  \end{center}
\end{figure}

\subsubsection*{Possible symmetry breaking away from charge neutrality}

We now turn our attention to the situation when the surface state is away from charge neutrality, but not completely filled/empty.
Generally strongly interacting materials display a whole range of different ground states as function of charge carrier density~\cite{Dagotto2005-bn}.
Three layer RG is no exception, both ferromagnetism~\cite{Zhou2021-wf} and superconductivity~\cite{Zhou2021-of} appear at various doping intervals.
It is only natural to assume that thicker RG samples should host similarly rich many-body phases.

Our mean-field calculations at slight doping (2.16$\times$10$^{12}$ cm$^{-2}$, Fig.~\ref{fig:ferro}b) suggest that the gap closes, but the surface magnetization does not vanish.
Instead, it forms a ferromagnetic pattern with a $(\sqrt{3} \times \sqrt{3})R30^{\circ}$ enlarged unit cell (Fig.~\ref{fig:ferro}c).
At this doping, mean-field calculations suffice, since we are dealing with a magnetic state.
Nevertheless, we have checked that DMRG calculations also reproduce this enlarged unit cell and the magnetizations agree well with the mean-field ones.
Interestingly, this translation symmetry breaking is also present in the total LDOS, which is the quantity that is measured by STM.

We have been able to identify an area of the sample with slight $p$ doping as evidenced by the spectrum shown in Fig.~\ref{fig:ferro}a.
Measuring the $dI/dV$ signal inside the surface state at 20 mV bias voltage, shows a $\sqrt{3} \times \sqrt{3}$ modulation which enlarges the unit cell, both in the graphite lattice and in the Fourier transform of the data (Fig.~\ref{fig:ferro}d).
The presence of this superlattice is not evidence in itself for the predicted ferromagnetic pattern, because such patterns can be also created by intervalley scattering due to lattice defects, such as vacancies.
To rule out such surface defects as the origin of the pattern, we show in Fig. S14 a 12$\times$12 nm topographic image showing that vacancies are not present on the top graphene layer around the area of interest.
Furthermore, strong lattice defects cause intervalley scattering of all states regardless of energy, albeit with decreasing amplitude~\cite{Deshpande2009-rv}.
In our case, measuring at a 100 mV bias voltage, outside the surface state, in the same area of the sample, simply shows the expected periodicity of graphite (Fig.~\ref{fig:ferro}e).
This is unusual behavior for a defect induced LDOS modulation, providing a possible signature for the predicted ferromagnetic state.
However, subsurface defects can also lead to a slight modulation of the LDOS in the top graphene layer~\cite{Dutreix2016-ye} and the behavior of these is less well studied.
Therefore we can not completely rule out the defect origin of the $\sqrt{3}$ modulation we observe.

Ferromagnetism has been predicted at finite doping~\cite{Olsen2013-xn} and recently measured in the compressibility of trilayer RG~\cite{Zhou2021-wf}.
Our calculations suggest that ferromagnetism is also present in thick RG, and is accompanied by translation symmetry breaking.
At present it is still unclear whether the interaction induced $\sqrt{3} \times \sqrt{3}$ instability is accompanied by the formation of a Kekul\'{e} gap~\cite{Weeks2010-pt,Roy2010-ju,Gamayun2018-cf}, or some other valley order~\cite{Zhang2011-wu,Szabo2022-vh}.

\section*{Discussion}

We have shown that interactions in the surface state of thick RG produce a ground state that is fundamentally different from the mean-field prediction, having a domain structure of gapped and gapless surface charge density.
According to our DMRG calculations this domain structure is a result of the  degenerate ground state.
The fundamental difference between 3 to 4 layers and thick samples is that in the latter the surface states on opposite graphene terminations do not overlap~\cite{Shi2020-bv}.
Such an overlap changes the nature of the quantum magnet and lifts the characteristic degeneracy, resulting in a completely gapped ground state, as recently shown in spin chains of triangulene~\cite{Mishra2021-mr}.
As a consequence, the surface charge density in tetra-layers is always gapped and magnetic, therefore it is well described by mean-field theory~\cite{Kerelsky2021-os}.
Perhaps most importantly, RG provides a platform for the straightforward tuning of the correlated phase discussed here, through electric and magnetic fields~\cite{Shi2020-bv} and by mechanical strain, in a simple defect free quantum material.

Although we do not directly measure the sample magnetization, charge transport measurements of Shi et al. show hysteretic behavior of the resistance in a magnetic field, supporting the notion that thick RG hosts a non-zero magnetic moment~\cite{Shi2020-bv}.
Their measurements also hinted at the formation of insulating domains, which our STM measurements confirm.
Since the interacting surface state is easily accessible, future studies by spin polarized STM would be especially fruitful in directly identifying the surface magnetic moments.

Our DMRG calculations show that at charge neutrality, the spin and charge interactions in RG result in a degenerate ground state and a bulk spin gap.
These properties are also shared by a simpler toy model, the 1D quantum spin chain with a bulk spin of 2~\cite{Haldane1983-sb,Schollwock1995-fl}, raising the question: what other properties of this famous toy model are shared by RG?
Further theoretical and experimental investigations exploring the properties of this new system hold the promise of revealing the existence of non-local topological order, spin or charge fractionalization~\cite{Schollwock1995-fl,Pollmann2012-kq,Mishra2021-mr} and perhaps unconventional superconductivity~\cite{Volovik2013-qx,Zhou2021-of}.


\subsubsection*{Materials and methods}

  Samples of RG were exfoliated onto Si/SiO$_2$ surfaces from natural graphite crystals ("graphenium"), purchased from NGS Trading \& Consulting GmbH.
  RG crystals were identified by Raman spectroscopy, with a laser excitation of 532 nm.
  Before STM measurements Ti/Au electrodes were evaporated onto the samples using a stencil mask.
  STM measurements were conducted in a commercial UHV STM (RHK Freedom), at a temperature of 9.6 K and a base pressure of 5$\times$10$^{-11}$ Torr.
  For variable temperature measurements, a resistive heater was used with a closed loop temperature controller.
  Mechanically cut Pt/Ir tips were calibrated on a Au (111) surface prior to measurement.

\newpage


\bibliographystyle{naturemag}

\subsubsection*{Acknowledgements}
  
  The work was conducted within the framework of the Topology in Nanomaterials Lendulet project, Grant No. LP2017-9/2/2017, with support from the European H2020 GrapheneCore3 Project No. 881603.
  LT acknowledges support from the Elvonal grant KKP 138144.
  TZ acknowledges support from the UNKP-20-4 New National Excellence Program of the Ministry for Innovation and Technology through NKFIH.
  PV and LO acknowledges the support of the Janos Bolyai Research Scholarship and LO for the Bolyai+ Scholarship of the Hungarian Academy of Sciences.
  JK and LO acknowledges the support of Ministry of Innovation and Technology of Hungary within the Quantum Information National Laboratory.

\subsubsection*{Author contributions}

  MSMI prepared and characterized the samples.
  STM measurements were performed by MSMI, KK and PNI.
  IH designed and performed the DMRG and mean-field calculations, ZT and PV provided the various tight-binding Hamiltonians.
  KM, PK and AP were involved in the Raman spectroscopic mapping and sample preparation.
  ZT performed the ab-initio calculations, under the supervision of JK with the assistance of PV.
  OL and AA performed the tight-binding modelling of quantum confinement effects.
  LT contributed to data analysis.
  PNI conceived and coordinated the project and wrote the manuscript with input from all authors.


\newpage


\begin{center}
  \section*{Supplementary information for: Observation of competing, correlated ground states in the flat band of rhombohedral graphite}
\end{center}

\renewcommand{\thesection}{S\arabic{section}}
\renewcommand{\thefigure}{S\arabic{figure}}
\setcounter{figure}{0}

\section{Further experimental details}
\label{sec:methods}

\subsection{Sample preparation}
\label{sec:sample_prep}
  
\begin{figure}[h]
    \includegraphics[width = 1 \textwidth]{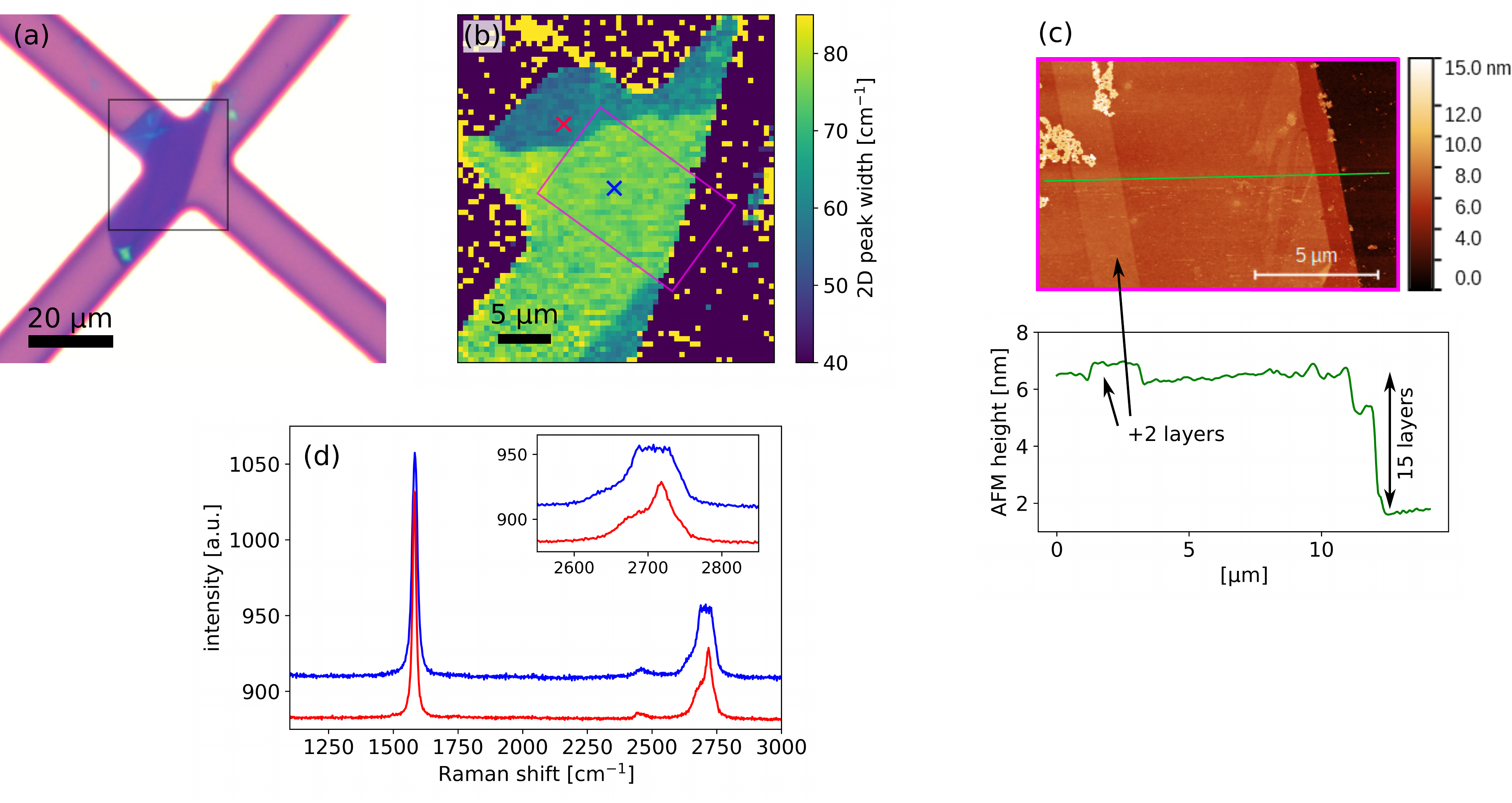}
    \caption{
      \textbf{Sample details.}
      \textbf{(a)} Optical microscope image of the RG flake, and the gold electrical contacts on a SiO$_2$ support.
      \textbf{(b)} Raman spectra map in the black rectangle shown in (a). The color scale shows the width of the 2D peak, resulting from Lorentzian fitting. The sample has a large rhombohedral area, where the width of the 2D peak is large. Blue and red crosses show the positions on the Raman spectra shown in (d). Purple rectangle shows the area of the AFM measurement in (c).
      \textbf{(c)} AFM measurement of the RG flake, showing its thickness of 15 graphene layers. The flake contains a ribbon-like area, with additional 2 graphene layers, all ABC stacked. Height section in green, measured across the green line.
      \textbf{(d)} Two Raman spectra from the map shown in (b), measured at the positions shown by the red and blue crosses. Spectra are shifted for clarity. Excitation wavelength was 532 nm.
    }
  \label{fig:sample}
  \end{figure}

  Thick RG flakes, with ABC type stacking have a very specific 2D peak profile, which can be used to distinguish them from Bernal stacked graphite \cite{Henni2016-yc}.
  An optical microscopy image of the sample discussed in the main text can be seen in Fig. \ref{fig:sample}a, supported on a Si/SiO$_2$ substrate and contacted by gold electrodes.
  The ABC stacked area of the sample was identified by Raman mapping.
  By plotting the width of the 2D peak across the sample surface, we can observe that most of the flake has rhombohedral stacking (see Fig. \ref{fig:sample}b).
  The thickness of the sample was determined by AFM measurements.
  The middle of the flake is made up of ABC stacked graphene layers, with a ribbon-like area, having two extra graphene layers of ABC stacking.
  An AFM image of the sample can be seen in Fig. \ref{fig:sample}c.

\newpage

\subsection{STM measurement details and tip calibration}

  STM measurements were conducted in a commercial UHV STM from RHK (RHK Freedom), at a temperature of 9K.
  The system allows for variable temperature measurements while keeping the tip engaged, we used this to characterize the same areas on the sample at increasing temperature, allowing us to determine the local critical temperature.
  For spectroscopy measurements, a Lock-In amplifier was used with a modulation frequency of 1372 Hz.
  The bias modulation amplitude, measured as the RMS of the signal, ranged from 2 mV for the highest energy resolution individual $dI/dV$ spectra to 10 mV, in the case of $dI/dV$ maps.

\begin{figure}[h]
    \includegraphics[width = 1 \textwidth]{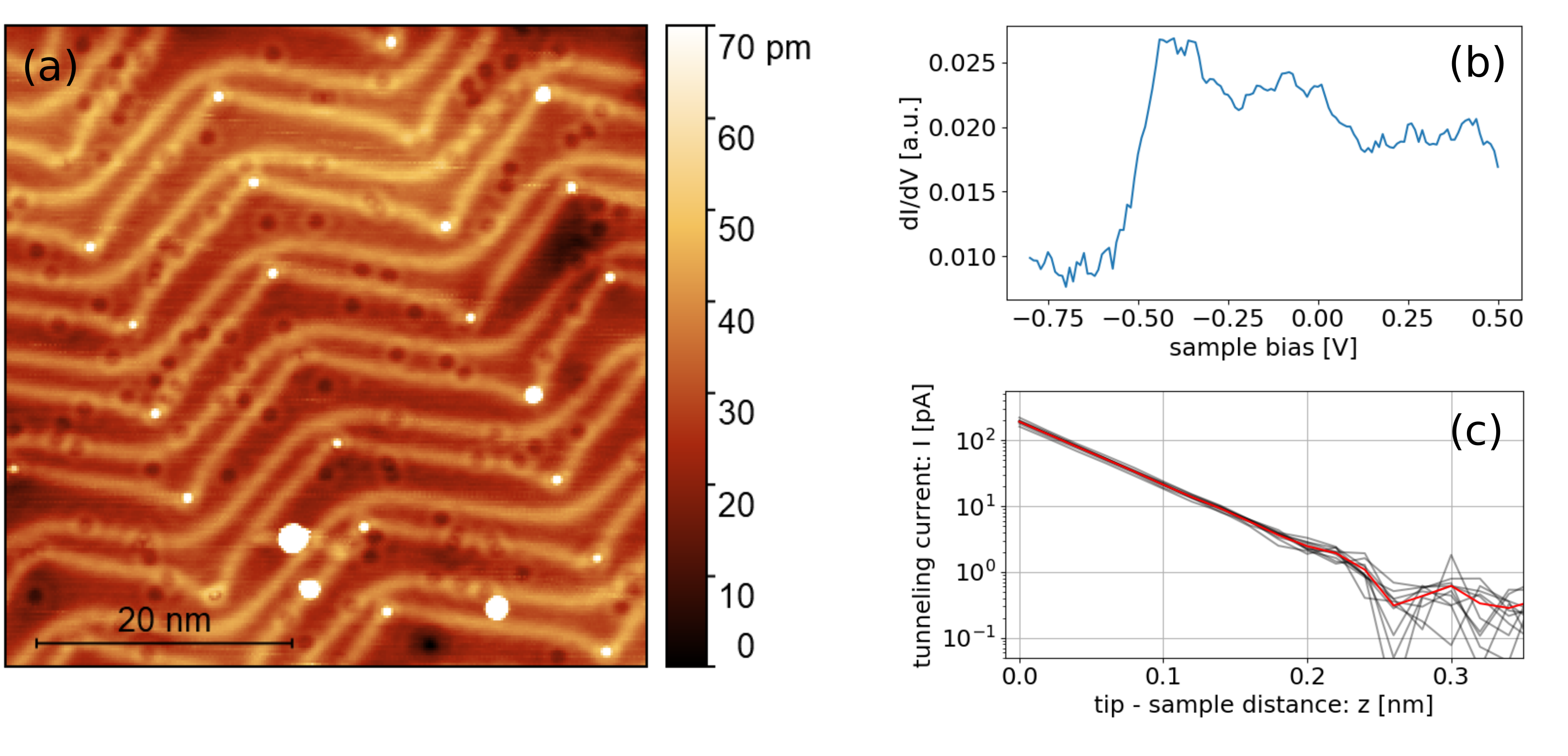}
    \caption{
      \textbf{STM tip calibration.}
      \textbf{(a)} STM topographic image of a gold (111) surface, showing the herringbone reconstruction.
      Circular rings are the LDOS modulation due to surface electron scattering and interference.
      Imaging parameters: 500 pA, 500 mV.
      \textbf{(b)} $dI/dV$ spectrum showing the gold (111) surface state.
      \textbf{(c)} Measurement of the tunneling current decay with 10 individual $I(z)$ curves (gray curves) measured on the gold surface.
      Red curve is an average of the displayed $I(z)$ spectra.
    }
  \label{fig:stmtip}
  \end{figure}

  For the STM measurements, sharp Pt/Ir tips were cut mechanically, allowing the optical positioning of the tip onto the RG flake.
  The STM tips were calibrated on a gold (111) surface.
  Tip were used on the RG samples, when they reproducibility imaged the herringbone structure of the gold and show the (111) surface state (see Fig.~\ref{fig:stmtip}).

\newpage

\section{Calculation details}
\label{sec:calc}

  \subsection{Density functional theory calculations}
  \label{sec:calc_dft}

    The optimized geometry and ground state Hamiltonian and overlap matrix elements of each structure were self consistently obtained by the SIESTA implementation of density functional theory (DFT) \cite{soler2002siesta,artacho2008siesta}.
    SIESTA employs norm-conserving pseudopotentials to account for the core electrons and linear combination of atomic orbitals to construct the valence states.
    For all cases the considered samples were separated with a minimum of 1.35 nm thick vacuum in the perpendicular direction.
    The generalized gradient approximation of the exchange and the correlation functional was used with Perdew–Burke–Ernzerhof parametrization \cite{perdew1996generalized} with a double-$\zeta$ polarized basis set.
    The geometry optimizations were performed until the forces were smaller than 0.1 eV nm$^{-1}$. 
    The geometry of the considered structures were optimized for every configuration, initiated from the experimental in-plane lattice constant $a=0.246\, \mathrm{nm}$ and out-of-plane lattice constant $c=0.670\, \mathrm{nm}$.
    During the geometry relaxation the real-space grid was defined with an equivalent energy cutoff of 400 Ry and the Brillouin zone integration was sampled by a 120$\times$120$\times$1 Monkhorst–Pack k-grid \cite{monkhorst1976special}.
    After relaxation a self-consistent single-point calculation was done with LDA+U approximation \cite{dudarev1998electron} with collinear spin.
    During the collinear single-point calculation the real-space mesh grid was increased to 1000 Ry.
    The effective Coulomb repulsion was chosen to 1.9 eV in order to match the magnetization that was calculated by Francesco Mauri et al. \cite{Pamuk2017-mj} using HSE hybrid functionals.
    The sisl tool (N. Papior, sisl: v0.10.0 (2020)) was used to extract the PDOS from SIESTA calculations, sampling the Brillouin zone with a 900$\times$900$\times$1 k-points set.
    
\subsection{Effect of mechanical strains}
\label{sec:theort-strains}
  
  Manufactured devices are commonly subject to mechanical distortions during the production, which can have a non-negligible impact on the electronic states and magnetic moments of the system.
  In order to investigate the effects we modeled in-plane and out-of-plane strains in our \textit{ab initio} calculations.
  We simulated the in-plane distortion by stretching the in-plane unit cell vectors along a carbon–carbon bond and allowing the atomic positions relax in the constrained unit cell.
  On the other hand out-of-plane strain was modeled by reducing the distance between graphene layers, without relaxation of the atomic positions.

\begin{figure}
    \begin{center}
    \includegraphics[width = 0.4 \textwidth]{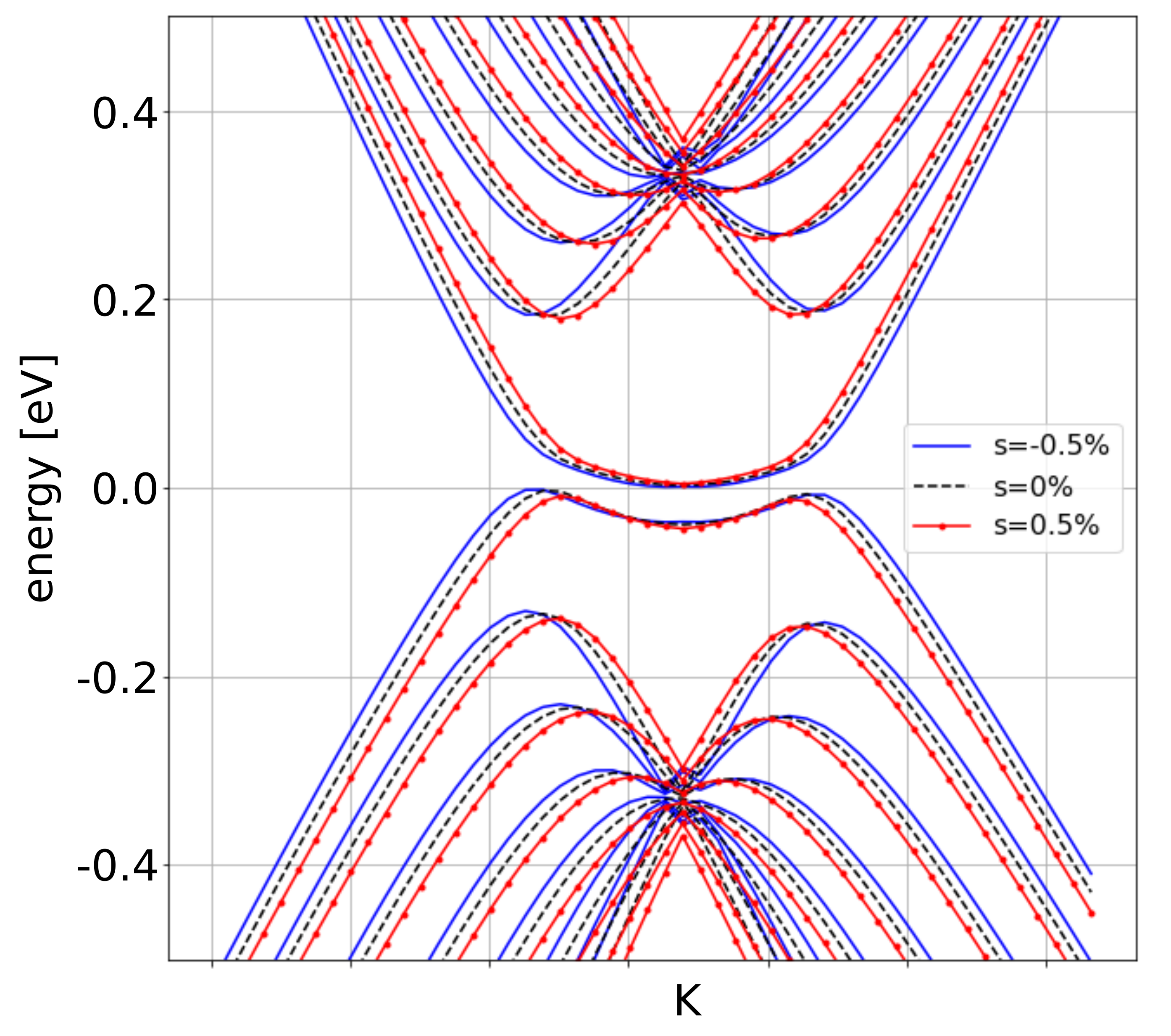}
    \caption{
      Electronic bands calculated using LDA+U approximation, showing the surface state splitting due to the N\'eel ground state (black dashed lines).
      Blue and red lines show the same calculation, but for $\pm$0.5\%\ in-plane strain.
      The surface bands don't shift in energy, as a function of in-plane or out-of-plane strain.
    }
    \label{fig:theory_strain_bands}
    \end{center}
    \end{figure}

  In Fig.~\ref{fig:theory_strain_bands} we show the bands for unstrained 8 layer RG, and the bands calculated for in-plane strains of $\pm$0.5\%.
  We chose 0.5\%\ strain because graphene and few layer graphene supported on SiO$_2$ shows typically strains in the range of $\pm$0.4\%.
  The calculations were carried out at the \emph{ab initio} level including interactions (see~\ref{sec:calc_dft}).
  The shift in energy of the bands is negligible both due to in-plane and out-of-plane strain.

\begin{figure}
    \begin{center}
    \includegraphics[width = 0.7 \textwidth]{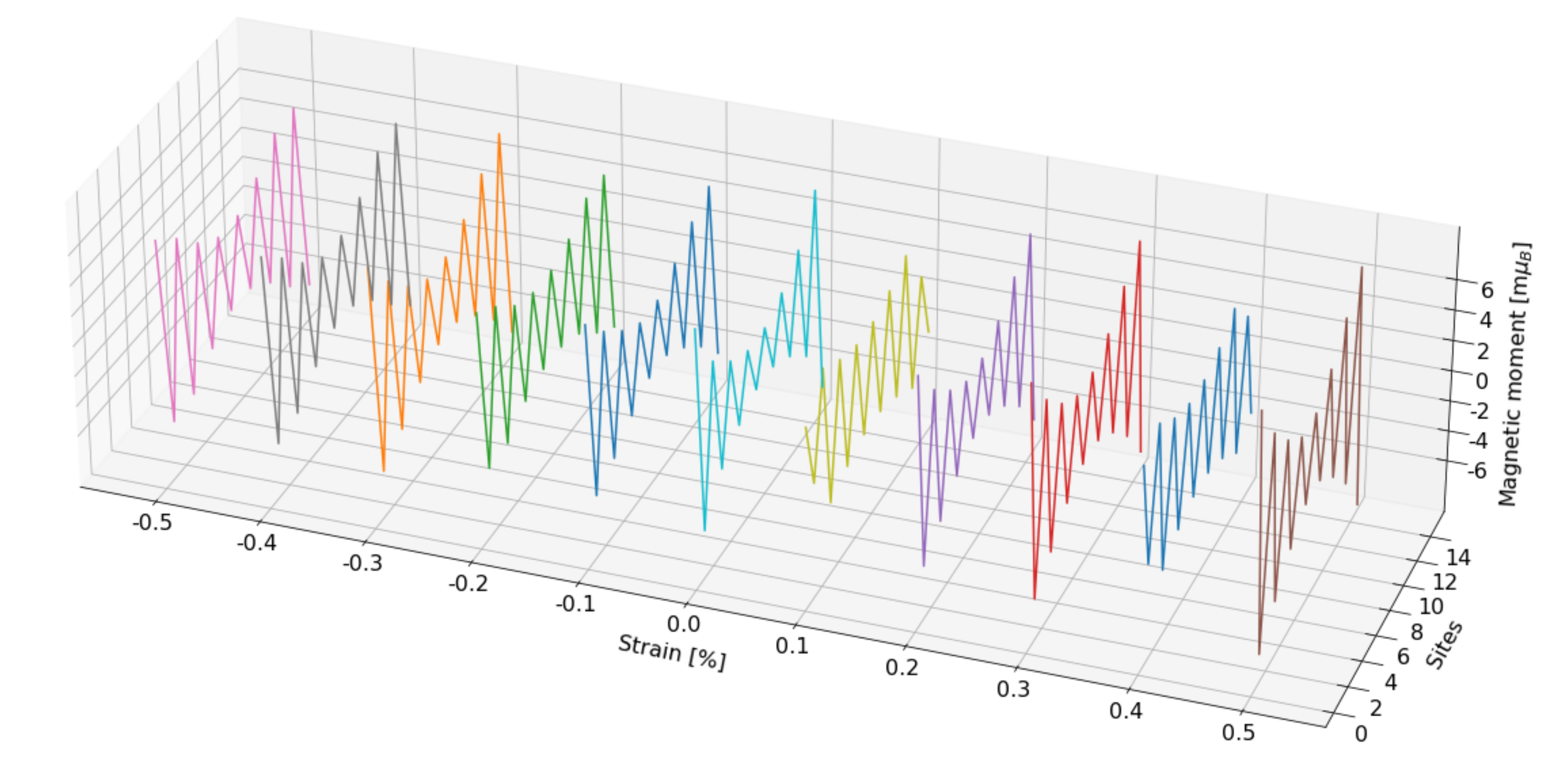}
    \caption{
      The impact of in-plane mechanical strain on the magnetic moments of an 8 layers thick sample. The individual, colored, solid lines show the magnetic moments on every site in $m \mu_B$ units.
    }
    \label{fig:theory_strain1}
    \end{center}
    \end{figure}

\begin{figure}
    \begin{center}
    \includegraphics[width = 0.7 \textwidth]{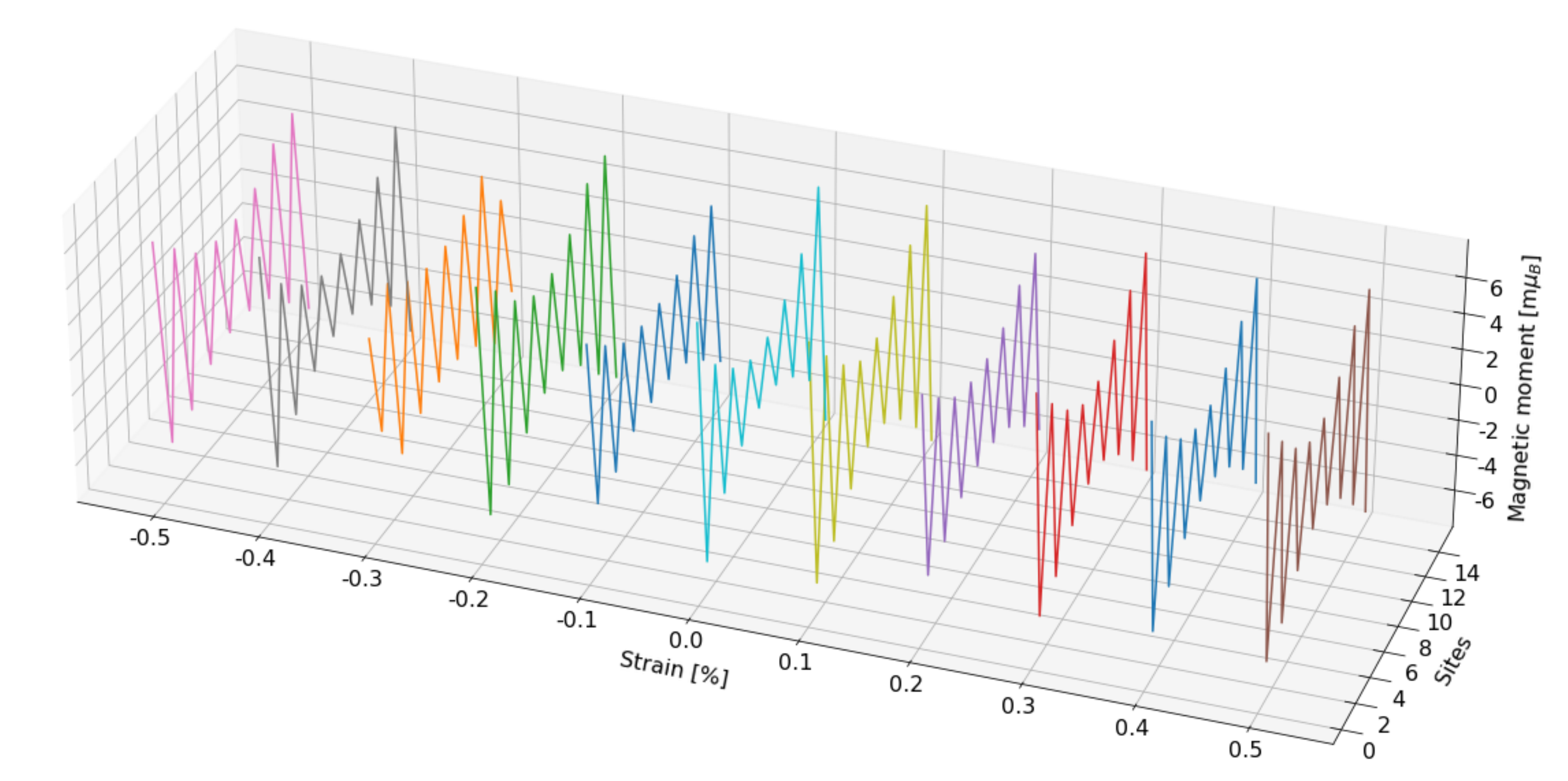}
    \caption{
      The impact of out-of-plane mechanical strain on the magnetic moments of an 8 layers thick sample. The individual, colored, solid lines show the magnetic moments on every site in $m \mu_B$ units.
    }
    \label{fig:theory_strain2}
    \end{center}
    \end{figure}

  We also calculated the magnetic moments of the 8 layers thick sample influenced by the distortions between $\pm 0.5 \%$. The results are shown on \ref{fig:theory_strain1} and \ref{fig:theory_strain2}.
  It can be seen, that the magnetic moments of the system do not alter significantly and they keep the decaying tendencies throughout the sample regardless of the type of the distortion.
  
  \label{sec:calc_strain}

  \subsection{Details of the density-matrix renormalization group calculations}
  \label{sec:dmrg}

    We use the SU(2) and U(1) variants of the DMRG algorithm \cite{hubig:_syten_toolk,hubig17:_symmet_protec_tensor_networ,hubig_2015,McCulloch_2007} to study the finite segments of RG systems.
    Although the method performs best for one-dimensional gapped systems, it has been successfully applied to study two-dimensional \cite{white_2d_dmrg} and three-dimensional systems very recently \cite{ummethum_numerics_2013,hagymasi_prl_2021}.
    In order to make the 2D RG system addressable for DMRG, we considered rectangular prism segments --  somewhat similar to the widely used cylindrical setups used for pure 2D models \cite{white_2d_dmrg} --, but with periodic boundaries in the $x-y$ (surface) planes, these ensure that the translation symmetry is not broken here and the surface effects can be properly simulated (see Fig. \ref{fig:theory_cell}).
    Due to the area-law scaling of the entanglement, the computation's cost increases exponentially with the linear sizes in the $x-y$ plane, while increasing the number of layers is cheap in comparison.
    The maximal 2D section we were able to handle with DMRG consists of $4\times 3=12$ sites (see Fig.~3c and 3d of the main text).
    We keep up to 12000 U(1) and 8000 SU(2) states in the corresponding simulations, such a high number of states is clearly desirable for a reliable extrapolation as it can be seen from Fig. 3b of the main text.
    In the U(1)-symmetric simulations, one cannot control the total spin of the state in advance but can be evaluated once the state is converged. 
    In addition, the degeneracy originating from the end spins makes the selection of a given spin state also difficult. To overcome this difficulty, the SU(2)-symmetric approach provides a clear solution, since it allows us to set the total spin of the state \emph{a priori} and to confirm the expected degeneracy occurring in an effective $S=2$ system.
    To extrapolate the energies to the error free limit, we used the two-site approximation of the full variance\cite{hubig_prb_2018}, since the evaluation of the latter quantity would be too expensive, on the other hand, it is an unbiased measure of the accuracy and not related to the optimization procedure like the truncation error in the two-site version of DMRG. 
    Due to its local-update nature, DMRG can get stuck in local minima, which can be avoided by using different initial states and pinning fields in the U(1)-symmetric case. 

    The DMRG method also allows us to test the reliability of the widely used mean-field approach (for more detail see the next section).
    Due to the limitation of the DMRG in the 2D plane, the comparison could be performed only for the $4\times 3$ geometry. To keep the computations relatively short, we considered 6-14 layers, thus 72-168 sites altogether.
    Due to the small 2D surface, the mean-field calculation gives the desired magnetic edge states only with smaller $U$ values, e.g. $U=0.5\gamma_0$ ($\gamma_0$ is the nearest neighbor hopping term), while for larger values, e.g. $U=\gamma_0$, it gives an antiferromagnetic ordering in the bulk as well, which is clearly an artifact of the mean-field approximation, since DMRG predicts the correct edge states but with a slower decay towards the bulk.
    It is also important to mention that the proper choice of $U$ is size-dependent, since for larger 2D segments the mean-field approach correctly predicts the edge states even for larger $U$ without a polarized bulk, in addition, in the infinite-size limit even $U=2\gamma_0$ does not result in spin-polarized bulk in agreement with the DFT calculations.
    All in all, regarding the $U$ region where the surface magnetization occurs in the mean-field calculation, we found an overall very good agreement between DMRG and the mean-field results for half filling and doping as well.
    \par As a closing remark we note that we only consider the magnetization originating from the spin degree of freedom, the orbital contributions are not taken into account.
    Thus, we call the $s_z = 0$ state a paramagnet, in experiments orbital contributions may overwhelm the paramagnetic response.

  \subsection{Details of the mean-field calculation}
  \label{sec:hubbard}

    We performed mean-field calculation for infinite as well as for finite segments of RG systems at finite and zero temperature. We briefly recall the main steps for the former case.
    The tight-binding Hamiltonian of RG with the local Hubbard interaction can be written as:
    \begin{equation}
     \mathcal{H}=\sum_{n\boldsymbol{k}\sigma}(\varepsilon_{n\boldsymbol{k}\sigma}-\mu)\hat{c}^{\dagger}
    _ { n  \boldsymbol{k}
    \sigma }\hat{c}^{\phantom\dagger}_{n\boldsymbol{k}
    \sigma }+U\sum_i\hat{n}_{i\uparrow}\hat{n}_{i\downarrow},
    \end{equation}
    where $\hat{c}^{\dagger}_{n\boldsymbol{k}\sigma}$
    ($\hat{c}^{\phantom\dagger}_{n\boldsymbol{k}\sigma}$) operators
    create (destroy) a particle with wave number $\boldsymbol{k}$ with spin $\sigma$ in band $n$, $\hat{n}_{i\sigma}$ are the particle-number operators and $\mu$ is the chemical potential.
    Neglecting the fluctuation terms, $(\hat{n}_{i\uparrow}-\langle\hat{n}_{i\uparrow}\rangle)(\hat{n}_{i\downarrow}-\langle\hat{n}_{i\downarrow}\rangle)$, we arrive at a single-particle problem that can be diagonalized and solved self-consinstently for the unknown chemical potential and $\langle\hat{n}_{i\sigma}\rangle$ densities. Once the largest difference between the consecutive electron densities becomes smaller than $10^{-5}$ we consider them as converged. The state with the lowest free energy is considered as the physical solution.
    \par
\begin{figure}
    \begin{center}
    \includegraphics[width = 0.5 \textwidth]{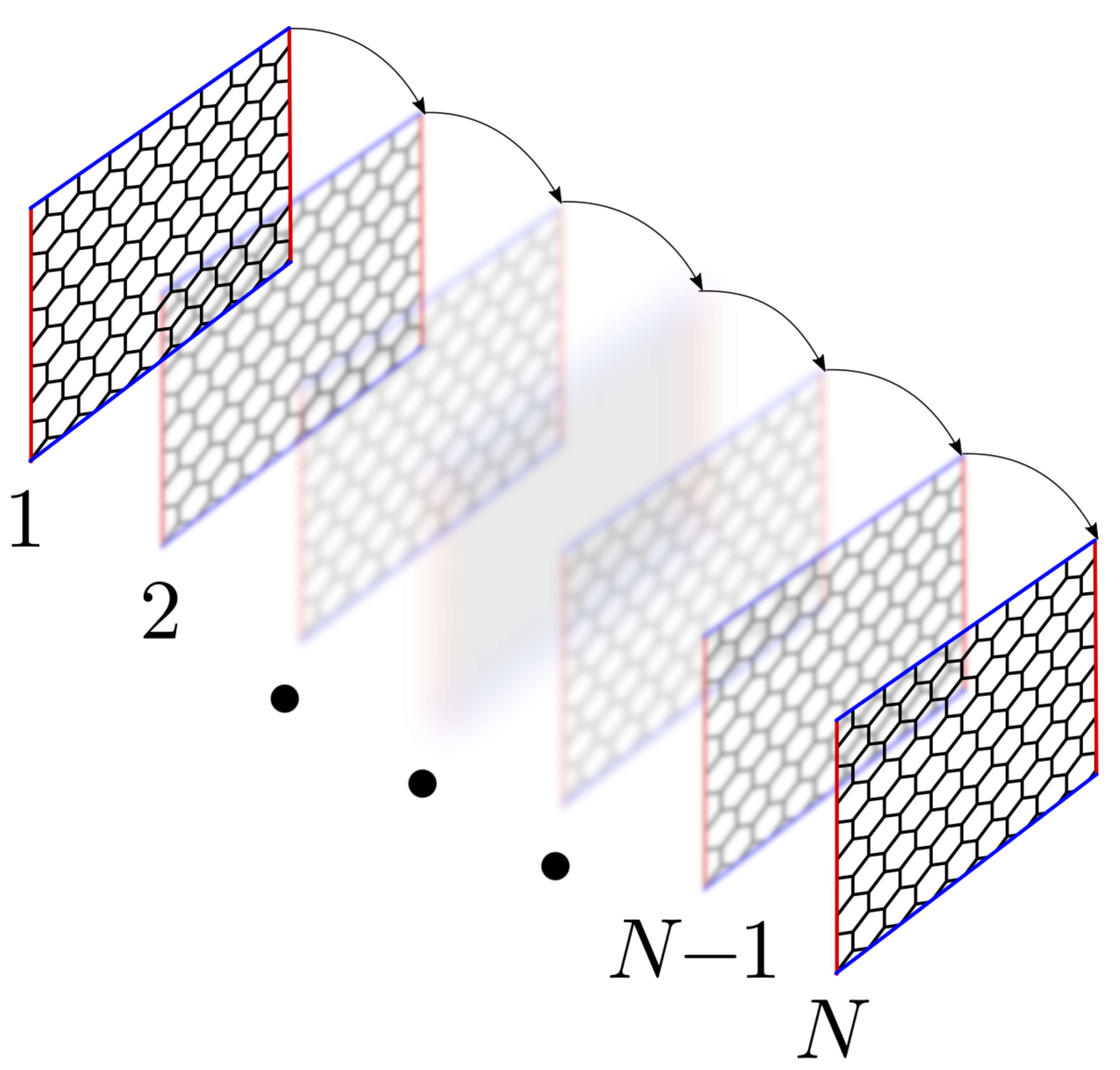}
    \caption{
      Finite calculation cell, with periodic boundary conditions along the blue and red edges. The interlayer hoppings are not shown for better visibility but are indicated by the arrows. We consider only the case with intralayer, nearest neighbor hopping, $\gamma_0 = 3$ eV, and interlayer hopping $0.1 \times \gamma_0$ in the mean-field and DMRG calculations.
    }
    \label{fig:theory_cell}
    \end{center}
    \end{figure}

 The advantage of the infinite system calculation is that we obtain results immediately in the thermodynamic limit.
    The integration over the Brillouin zone becomes very expensive if one wishes to consider larger unit cells that can host more complex magnetic patterns.
    To this end we reverted back to the simulation of finite systems, but with periodic boundaries in the $x-y$ directions (see Fig. \ref{fig:theory_cell}).
    Not only can we compare the mean-field magnetizations directly to the DMRG results, but we can consider much larger finite systems with mean-field as well. These can provide us quantitative insight into the doping effects.

\newpage

\section{STM measurements at varying sample thickness}
\label{sec:thickness}

  Due to the variation in the sample thickness in the rhombohedral part of the sample (see Fig. \ref{fig:sample}b,c), we were able able to check, in the same sample, how the surface state changes with increasing the RG thickness by two extra graphene layers.
  During our STM measurements we have identified the two layer thick step in the sample (see Fig. \ref{sec:thickness}a), also shown in the AFM image of Fig. \ref{fig:sample}c.
  We have performed multiple $dI/dV$ maps of the step area.
  Using a 32$\times$32 map measured in the area shown in Fig. \ref{sec:thickness}a, we plot the splitting values on the thicker and thinner part of the sample, respectively (see Fig. \ref{sec:thickness}b).
  Increasing the thickness by two graphene layers increases the mean splitting observed by 1.7 mV, as we can observe in the splitting histogram of Fig. \ref{sec:thickness}b.
  However, the standard deviations of the splitting values are somewhat larger, making the increase insignificant.
  Measurements on BN supported samples, with lower disorder will be needed to characterize the changes in mean splitting.
  The splitting was determined, as described in~\ref{sec:fitting} by fitting two Gaussians to the surface state $dI/dV$ peak.
  
\begin{figure}
  \begin{center}
    \includegraphics[width = 0.9 \textwidth]{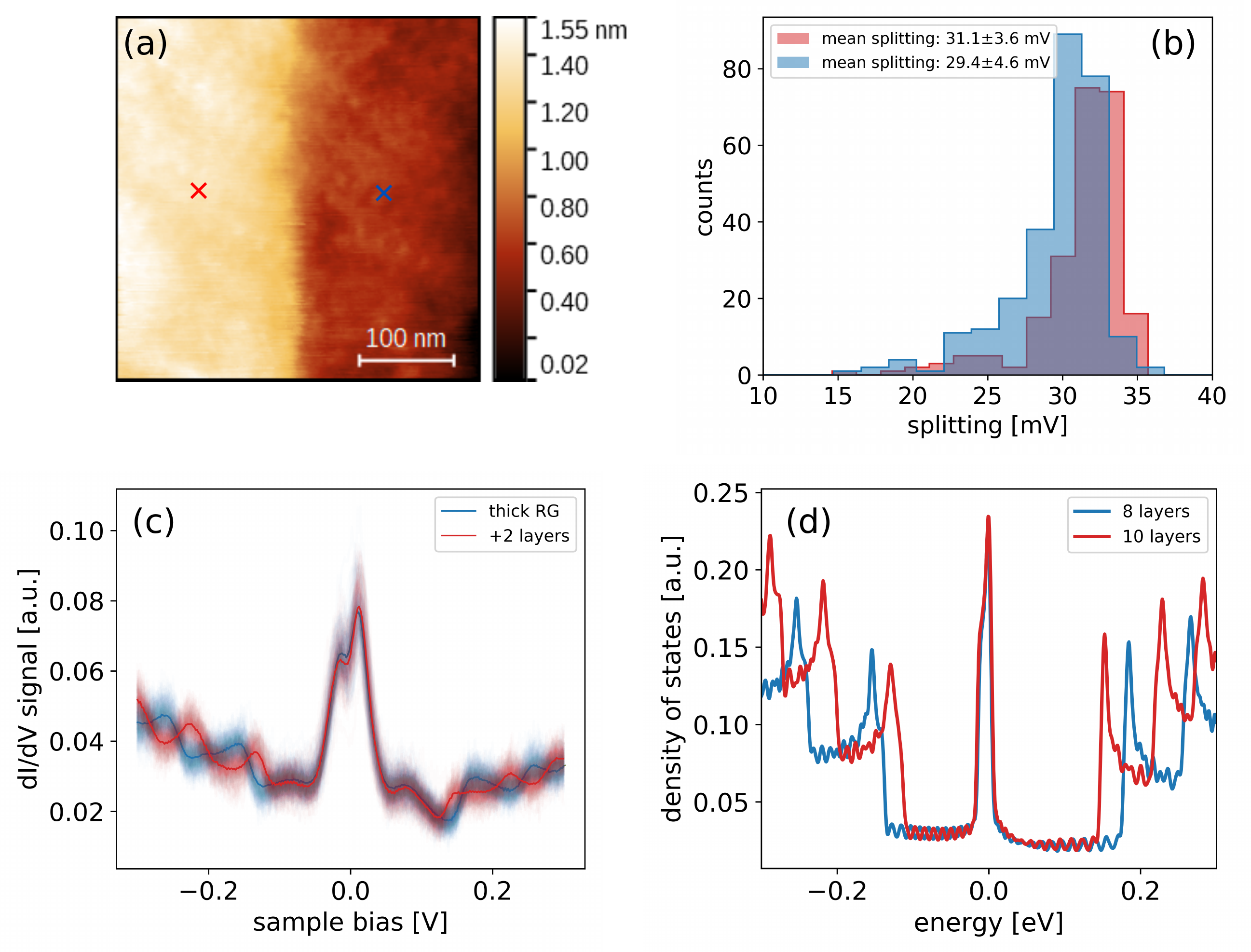}
    \caption{
      \textbf{Thickness dependence.}
      \textbf{(a)} STM topographic image of the step in the sample, also shown in Fig. \ref{fig:sample}a. Blue and red crosses show the positions of the $dI/dV$ spectra, shown in (c).
      \textbf{(b)} Blue and red histogram data are measured on the thin and thick part respectively.
      The splitting values result from a 32$\times$32 $dI/dV$ map measured in the area shown in (a).
      Values after the mean are the standard deviation of the splittings.
      \textbf{(c)} Semi transparent plots of 100 $dI/dV$ spectra on the thicker (red) and thinner (blue) part of the sample. Thicker blue and red spectra show the average on the two sides of the step.
      \textbf{(d)} Calculated (SIESTA) total density of states for an 8 and a 10 layer RG crystal.
    }
  \end{center}
  \label{fig:thickness}
  \end{figure}

  Selected spectra from the $dI/dV$ map can be seen in Fig. \ref{fig:thickness}c.
  Here we can observe that the largest difference in the spectra is not in the surface state peak itself, but the shifting of the steps above $\pm$0.1 V to lower voltages, with increasing sample thickness.
  The peaks are due to bulk bands of the RG sample and we observe a shift to lower energies by $\sim$20 meV on the thicker part of the sample.
  The measured spectra are reproduced well by calculating the total DOS of a RG slab of varying thickness, using DFT (see Fig. \ref{fig:thickness}d).
  The calculated spectra shown here do not contain many-body effects, thus the surface state shows no significant change with increasing thickness.
  However, the DOS peaks of the bulk bands show a shift to lower energies as we increase the thickness by two layers.
  In our calculations, the addition of two extra graphene layers shifts the DOS peaks by 20 meV, similarly to the measured shift.

  Interestingly, the experimental $dI/dV$ signal is best reproduced by a RG slab of 8 layer thickness, instead of a 15 layer one.
  So, there is a discrepancy between the thickness measured by AFM and the one estimated using the bulk band positions.
  A possible explanation could be that deep in the RG crystal, there is a stacking fault of the graphene layers.
  This would result in an ABC stacked crystal of smaller "effective" thickness, than the 15 layer determined by AFM.
  Based on our measurements on other RG crystals, it is unlikely that the stacking fault is near the "top" surface measured by STM, since in that case, the surface state can not be observed.
  The most likely scenario is that 8 layers deep from the top surface, there is a stacking fault that disrupts the SSH chain \cite{Su1979-ok,Slizovskiy2019-un} of the ABC stacking, giving rise to the "effective" thickness of 8 layers.
  We mention that in the case of 8 layers, the bottom and top surface states are still well separated electronically, with the overlap being effectively zero between them.
  This is also apparent in Fig. 3a of the main text, where we show the decay of the surface state wave-function into the bulk.

  We have measured a separate 14 and 17 layer RG sample, that show similar insulating and metallic domains across their surface.

\newpage

\section{Details of the Gaussian fitting}
\label{sec:fitting}

\begin{figure}[h]
  \begin{center}
  \includegraphics[width = 0.7 \textwidth]{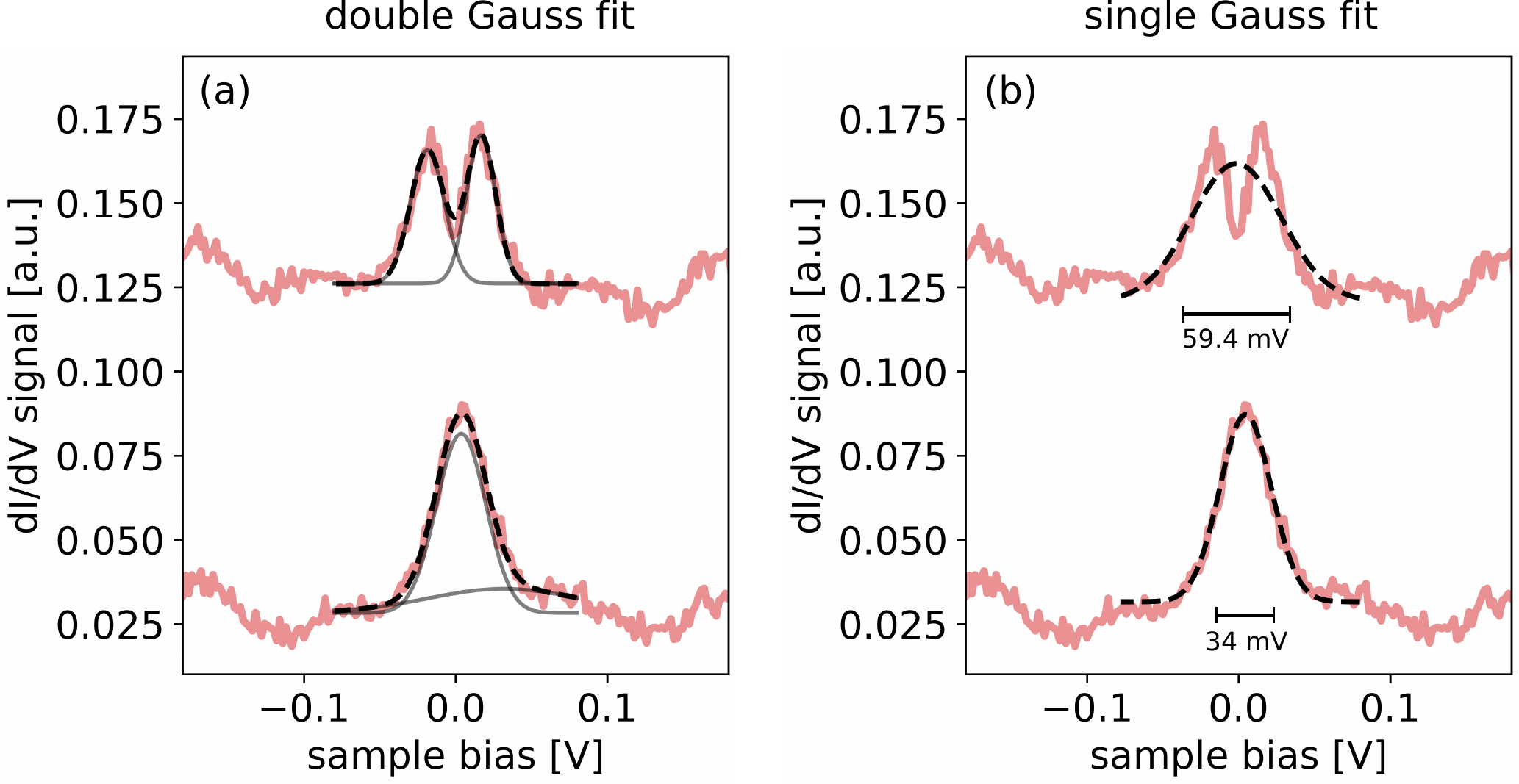}
    \caption{
      \textbf{Details of the Gaussian fitting.}
      \textbf{(a)} Double Gaussian fit to a $dI/dV$ spectrum showing a splitting of 35.3 mV and to one showing no splitting.
      \textbf{(b)} Single Gaussian fit to the same spectra shown in (a).
      FWHM ($w$) of the Gaussian fit is shown by the horizontal bar.
    }
  \label{fig:fitting}
  \end{center}
  \end{figure}

  To quantify the splitting of the surface state we fitted two Gaussian functions to the surface state peak.
  The absolute value of the voltage difference between the centers of these two Gaussians is defined to be the splitting of the surface state.
  Examples of the fits can be seen in Fig \ref{fig:fitting}a and in Fig. \ref{fig:fig2additional}.
  The fitting is done in the -80 and 80 mV interval around 0 bias, as shown in Fig. \ref{fig:fitting}.
  For the gapless state, described by a non-split peak, the shape can be fitted well by a single Gaussian curve.

  By fitting a double Gaussian we obtain the splitting maps shown in the main text and the supplement.
  We consider the peak to be described by a single Gaussian if one of the Gaussians has an amplitude 10\%\ less than the amplitude of the other Gaussian, or if the center of both Gaussians is on the same side of the Fermi level.
  In this case we set the splitting value to zero (dark areas in the splitting maps).
  One example of this can be seen in the bottom spectrum in Fig. \ref{fig:fitting}a.
  
  To have a common comparison between the gapped and gapless phases, we also use the FWHM ($w$) of the surface state LDOS peak, as measured by fitting a single Gaussian.
  For examples of this see Fig. \ref{fig:fitting}b.
  Here, the peaks with splitting will have FWHM values between 45 and 60 mV, with gapless peaks having a FWHM below 45 mV.

\newpage

\section{Extended data for Figure 2}
\label{sec:extended_fig2}

\subsection{Characterizing the local doping}
\label{sec:doping}

  To show that the area in Fig. 2 of the main text is mostly charge neutral, we quantify the local doping of RG in this region.
  This is achieved by measuring how much of the spectral weight of the peak is located below the Fermi level (zero bias).
  We define the \emph{fill ratio} of the peak as the area under the Fermi level divided by the total area of the peak in the voltage interval of $\pm$0.1 V, as also described in the main text.
  For a fill ratio of 0\%, the whole surface state is completely empty of electrons, while for a ratio of 100\%, the state is completely filled (see for example Fig. 1e of the main text).
  Therefore, a fill ratio of 50\%\ corresponds to the charge neutral surface state.

\begin{figure}
  \begin{center}
  \includegraphics[width = 0.7 \textwidth]{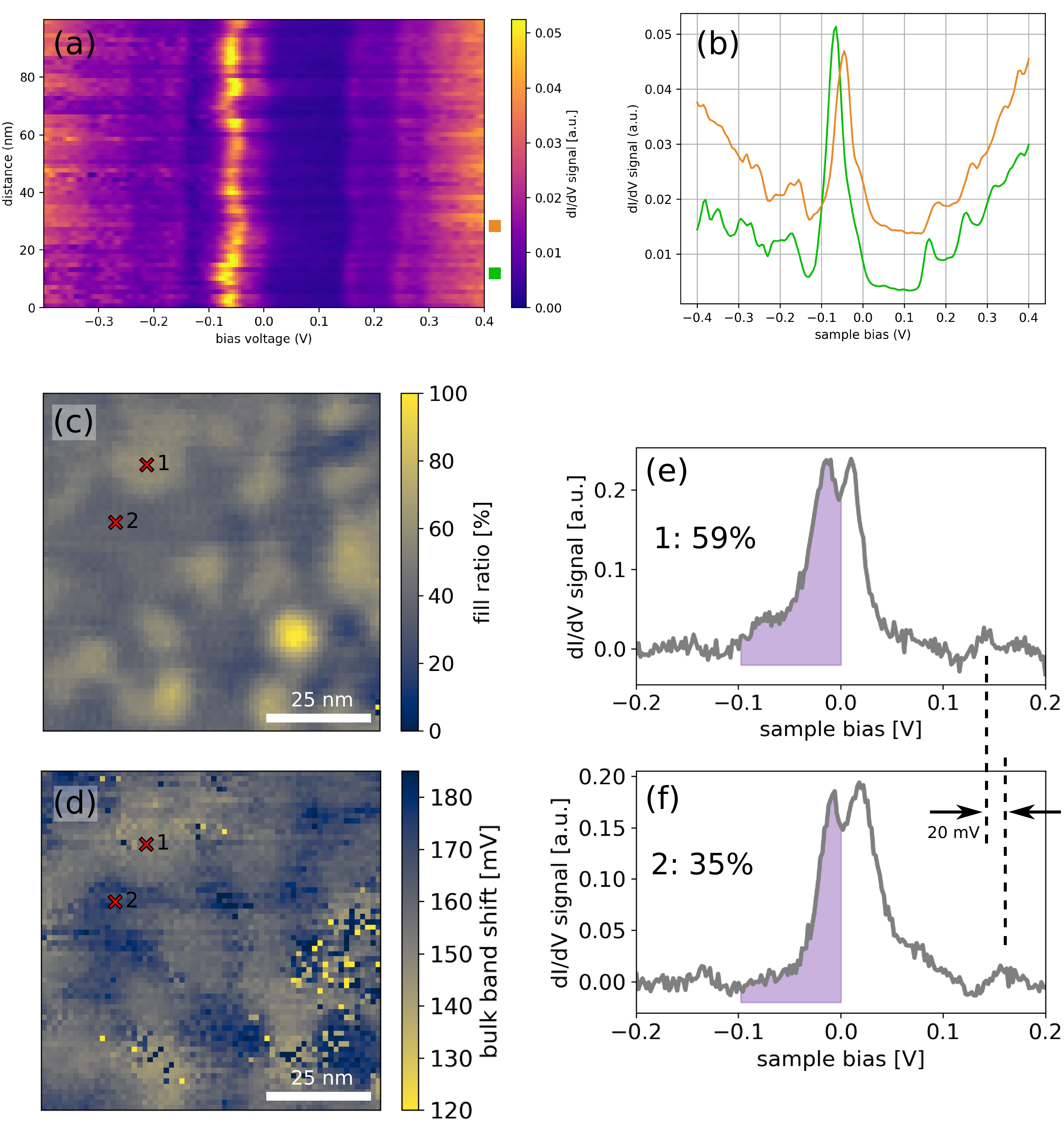}
    \caption{
      \textbf{Characterizing doping for the data in Fig. 2.}
      \textbf{(a)} $dI/dV$ spectra along a line on the sample, in an area where the surface state is completely filled.
      The bulk bands form peaked steps in the $dI/dV$ signal above 150 mV and below -100 mV.
      Selected spectra in (b) are marked by rectangles.
      \textbf{(b)} Selected spectra from (a), marked by the correspondingly colored rectangles.
      \textbf{(c)} Fill ratio of the surface state peak. Same data as in Fig. 2e of the main text, but plotted with a color scale matching the bulk band shift shown in (d).
      \textbf{(d)} Map of the first bulk band position in bias voltage across the same area.
      Outlying pixels are spectra where the fit has failed to converge, due to the lower signal to noise ratio of the bulk band peak.
      \textbf{(e, f)} Selected spectra from the map, showing the local fill ratio.
      The position of the first bulk band is marked by dashed lines.
      The shape of the surface state peak changes slightly, but its overall shift in energy is comparable to the shift of the bulk band.
    }
  \label{fig:fig2doping}
  \end{center}
  \end{figure}

  When the density of states at the Fermi level is negligible, the surface state peak shifts in energy along with other features of the LDOS, such as the peaks of the bulk bands.
  These bands are characteristic of the RG thickness and appear at energies higher than $\pm$100 meV (see also~\ref{sec:thickness} and Fig 1b of the main text).
  We show an example of the behavior of a completely filled surface state peak along a 100 nm line on the sample (Fig.~\ref{fig:fig2doping}a).
  Both the contour map and selected spectra show that the peak and the bulk bands shift in unison with the changing local doping.

  When the surface state is partially filled, the density of states at the Fermi levels is large and the local electrostatic disorder potential is not as effective in shifting the band energies. 
  In this situation many-body effects also play a role in determining the charge density at the Fermi level.
  For example, near 50\%\ fill ratio, the surface state peak shape can change, which changes the charge density at the Fermi level (splitting).
  This results in a diminished correlation of the bulk band position with the fill ratio (Fig.~\ref{fig:fig2doping}c, d).
  This makes the fill ratio a better gauge of the local charge density than the shift of the bulk bands.

  We have analyzed the local doping of the data presented in Fig. 2 of the main text, using the shift in energy of the first bulk band around 150 mV.
  To characterize the position in mV of the bulk band, we first subtract the background of the $dI/dV$ signal and fit a Gaussian to the first bulk band peak.
  In Fig.~\ref{fig:fig2doping}d we plot the shift of the peak on a color scale, in order to compare with the map of the fill ratio Fig.~\ref{fig:fig2doping}c.
  The two maps show a correlation, but the map of the fill ratio, shows more structure.
  We believe this is due to the many-body effects, which can change the shape of the surface state peak.


\subsection{Additional splitting maps}

\begin{figure}
  \begin{center}
  \includegraphics[width = 0.7 \textwidth]{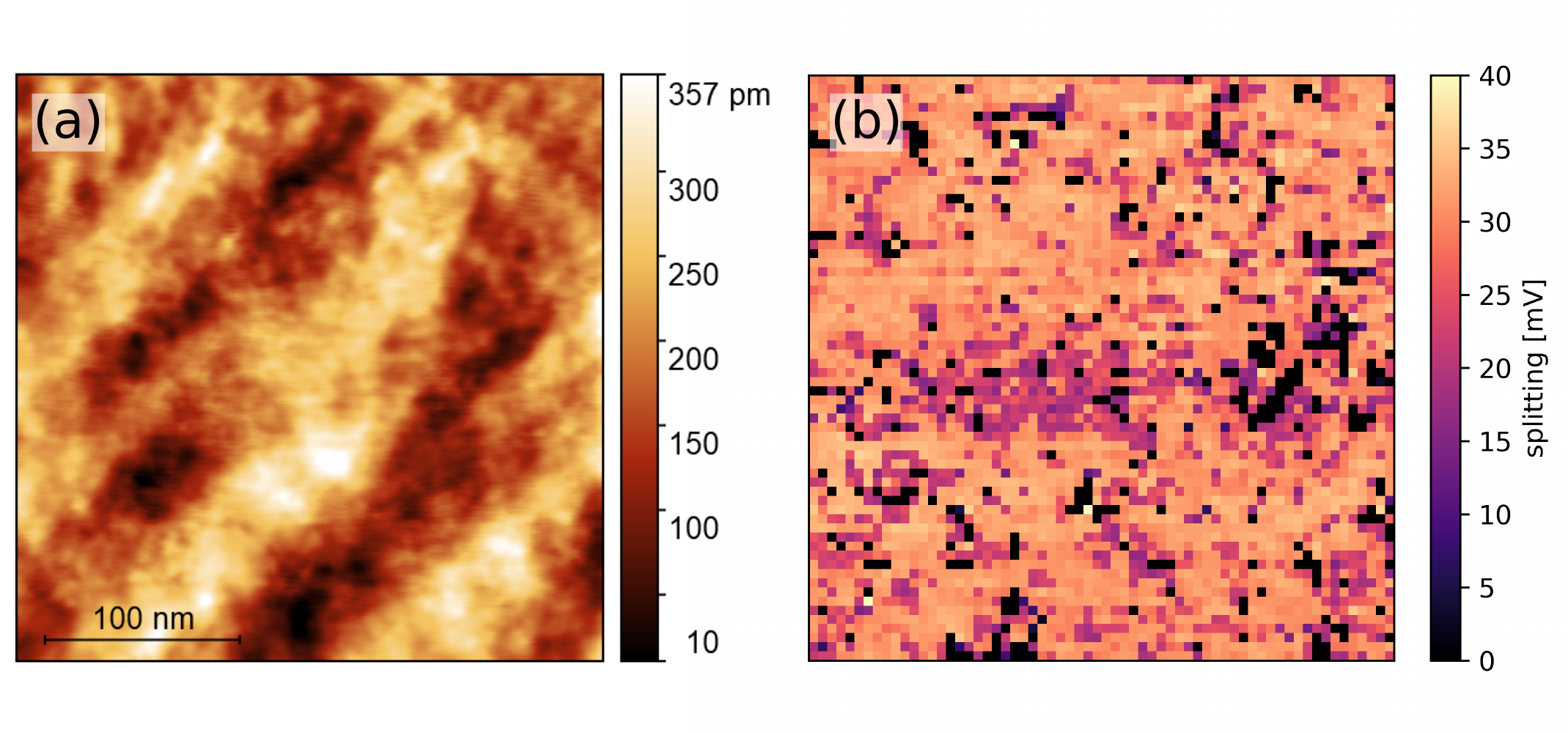}
    \caption{
      \textbf{Large area map.}
      \textbf{(a)} STM topography of a 300$\times$300 nm area of the sample presented in the main text (8 layers).
      \textbf{(b)} Map of the local splitting in the area shown in (a).
      The image is generated from a $dI/dV$ map with 64$\times$64 individual spectra.
      Stabilization parameters: 100 pA, 500 mV.
    }
  \label{fig:fig2large}
  \end{center}
  \end{figure}

\begin{figure}
  \begin{center}
  \includegraphics[width = 0.7 \textwidth]{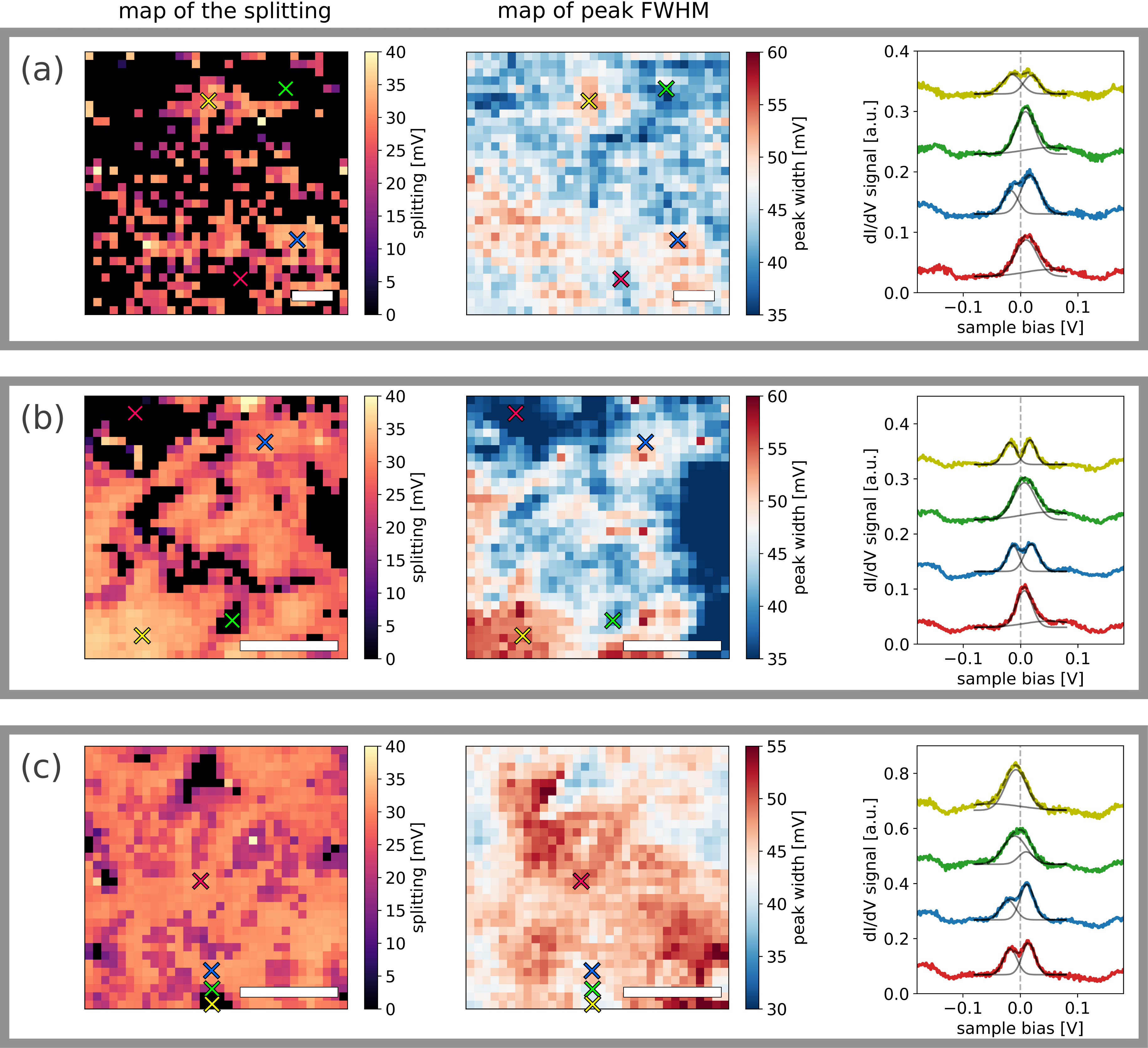}
    \caption{
      \textbf{Additional maps of the domain structure.}
      Splitting, surface state FWHM and selected spectra.
      Data in panels \textbf{(a)}, \textbf{(b)} and \textbf{(c)} are measured on different areas of the 8 layer sample.
      Left column: map of the splitting.
      Middle column: map of the FWHM.
      Right column: individual spectra in the areas marked by colored crosses in the maps.
      Individual Gaussian fits to the surface state peak are shown in gray, with the sum of the two Gaussians shown as a dashed gray line.
      White scale bars: 30 nm.
      Measured at a temperature of 9.6 K.
      Stabilization parameters for all three maps: 100 pA, 500 mV.
    }
  \label{fig:fig2additional}
  \end{center}
  \end{figure}

  In Fig.~\ref{fig:fig2large} and~\ref{fig:fig2additional} we show additional maps of the domain structure that forms between the gapped and gapless states, in the sample presented in the main text.
  Fig.~\ref{fig:fig2large} shows the splitting in a 300$\times$300 nm area.
  The maps shows a gapped surface state, with interspersed gapless regions.
  In Fig.~\ref{fig:fig2additional} we show higher resolution $dI/dV$ maps of the domain structure.
  All three maps are measured on different areas of the sample discussed in the main text.
  For each area we present the local splitting and the FWHM of the surface state peak.
  Selected spectra in the rightmost column also show some example individual spectra, with the Gaussian fits shown as black curves.

\newpage

\section{Extended data for Figure 4}
\label{sec:extended_fig4}

\begin{figure}[h]
  \begin{center}
  \includegraphics[width = 0.8 \textwidth]{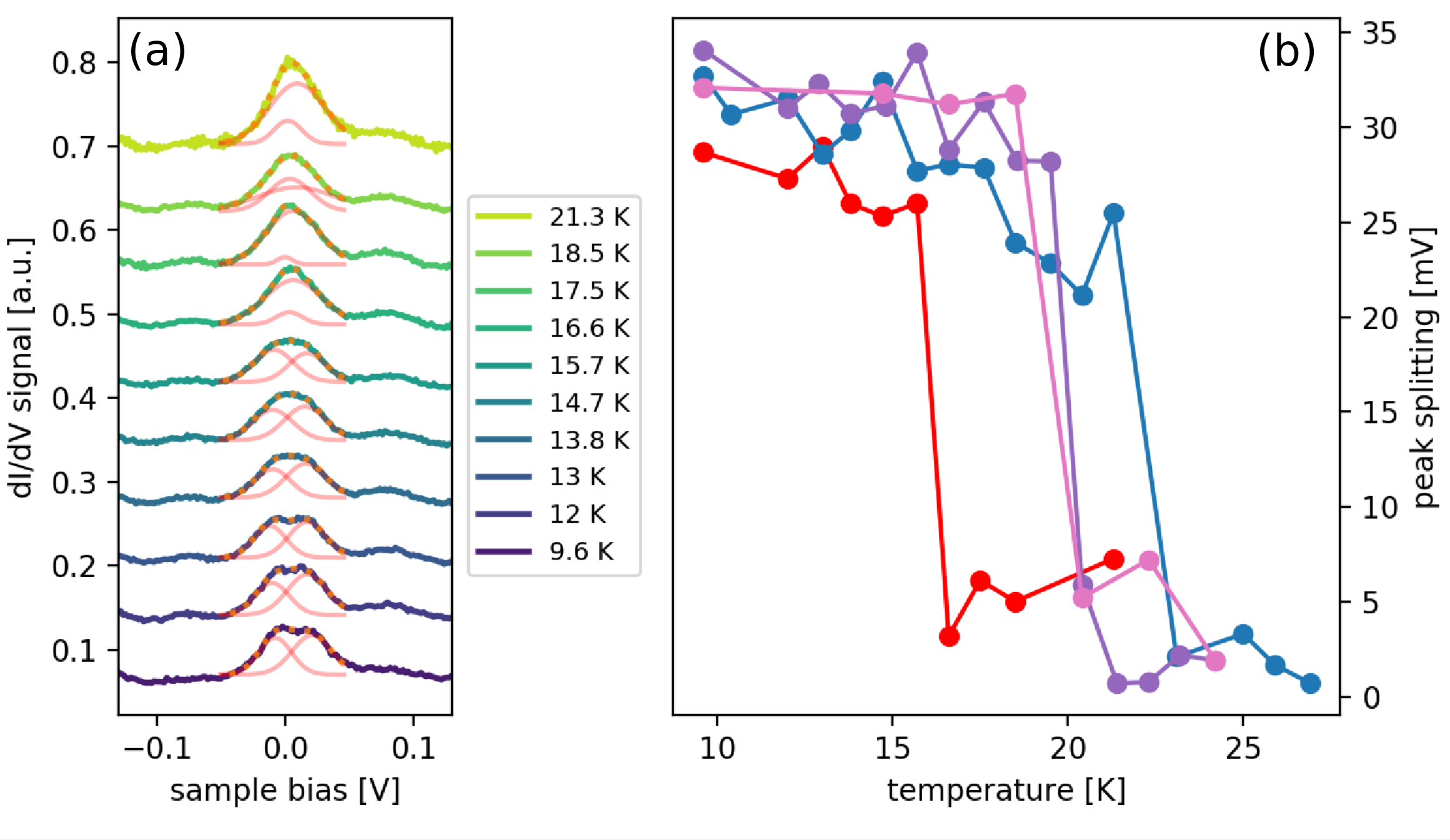}
    \caption{
      \textbf{Additional critical temperature measurements.}
      \textbf{(a)} $dI/dV$ spectra measured at increasing temperature in the same position on the sample.
      These spectra are used to plot the splitting and FWHM traces in Figure 4d of the main text.
      Double Gaussian fits plotted using orange curves.
      The sum of the Gaussians is plotted as an orange dotted line.
      \textbf{(b)} Red data points: the data shown in Figure 4d of the main text.
      Blue, magenta and purple data points: Critical temperature measurements at three additional areas of the sample.
    }
  \label{fig:fig4extended2}
  \end{center}
  \end{figure}

  In Fig. \ref{fig:fig4extended2}a we plot the $dI/dV$ spectra used to obtain the splitting and FWHM versus temperature plot in Fig. 4d of the main text.
  Fits to the surface state peak are plotted as orange Gaussians.
  In Fig. \ref{fig:fig4extended2}b we show data from three additional areas of the sample surface (blue, purple, magenta data points).
  The red data points are the ones shown in the main text, Figure 4d.
  The $T_\mathrm{C}$ differs slightly from one area to another.
  The highest value of the critical temperature we have measured is 22 K.
  \\

\begin{figure}
  \begin{center}
  \includegraphics[width = 0.9 \textwidth]{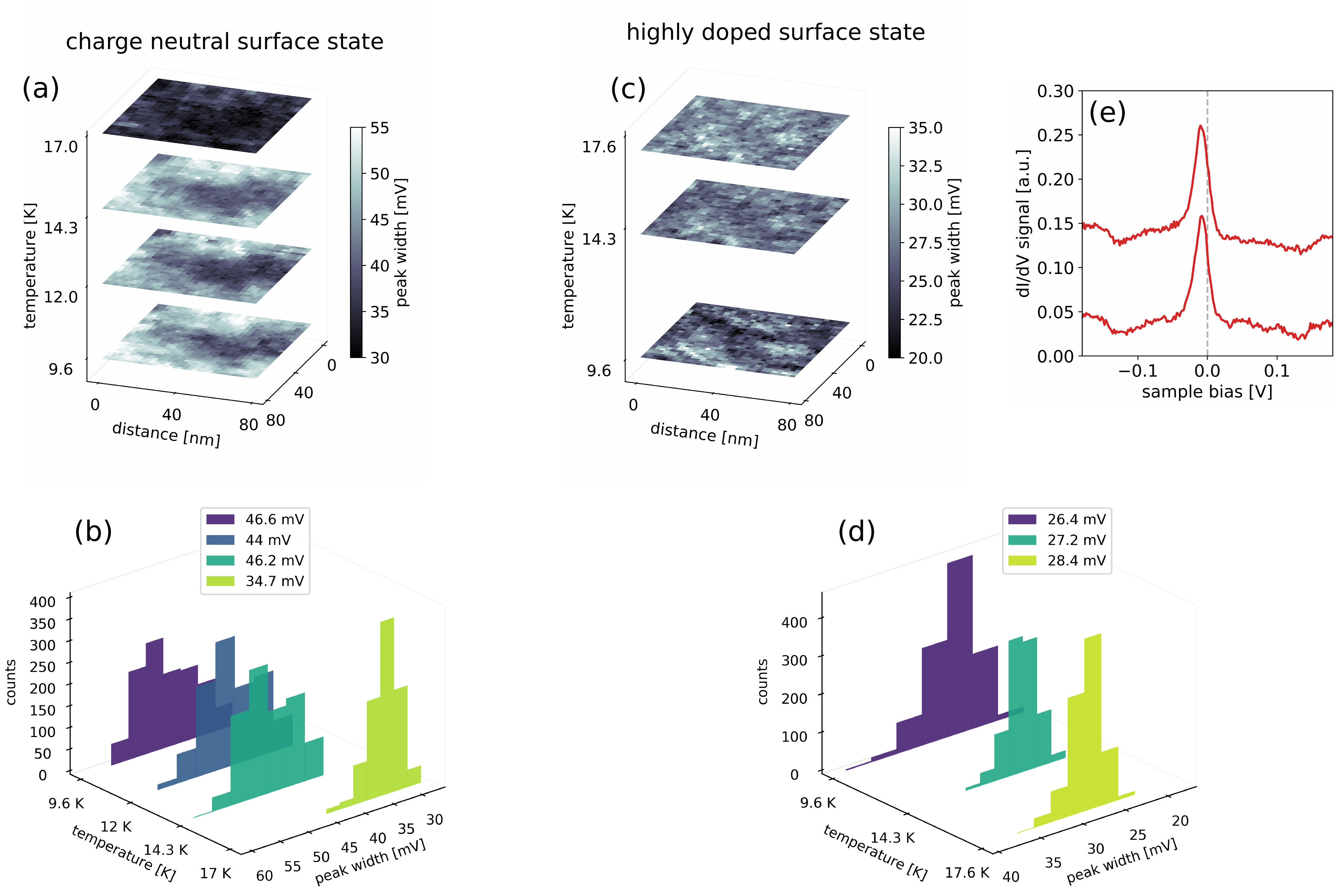}
    \caption{
      \textbf{Temperature dependence of the strongly doped surface state.}
      \textbf{(a)} Maps of the surface state FWHM at increasing temperature.
      The same data as shown in Fig. 4b of the main text.
      \textbf{(b)} Histogram of the FWHM values at increasing temperatures for the data shown in (a).
      Mean FWHM values shown in the inset.
      \textbf{(c)} FWHM maps at increasing temperature for an area of the sample where the surface state is highly $n$ doped.
      \textbf{(d)} Histogram of the FWHM values at increasing temperatures for the data shown in (c).
      Mean FWHM values shown in the inset.
      \textbf{(e)} Selected spectra from (c), measured at 9K.
    }
  \label{fig:fig4extended}
  \end{center}
  \end{figure}

  We also compare the temperature dependence of the half filled surface state to the almost completely $n$ doped case.
  Although the smallest scale for the doping inhomogeneity is $\sim$20 nm~\cite{Zhang2009-gn}, by moving the measurement area by $\sim$1 $\mu$m or more, we are able to find areas where the doping is further away from charge neutrality over the whole measurement area of $80 \times 80$ nm.
  Such an area is shown in Fig.~\ref{fig:fig4extended}c.

  In Fig. \ref{fig:fig4extended}a,b, we show the FWHM of maps of the area presented in Fig. 4b of the main text.
  This area shows an abrupt decrease in the FWHM over the whole $80 \times 80$ nm surface above a temperature of 16.6 K.
  The mean value of the peak width decreases by more then 10 meV, from 14.3 K to 17 K.
  When performing the same temperature dependent measurement on an area of the sample where the surface state is strongly $n$ doped (Fig.~\ref{fig:fig4extended}c), we observe no such decrease in the mean peak width.
  Instead the the mean peak FWHM increases by 2 meV, which is the amount expected from the increased broadening due to the increased temperature.

\newpage

\section{Extended data for Figure 5}
\label{sec:fig5extended}

\begin{figure}[h]
  \begin{center}
  \includegraphics[width = 0.8 \textwidth]{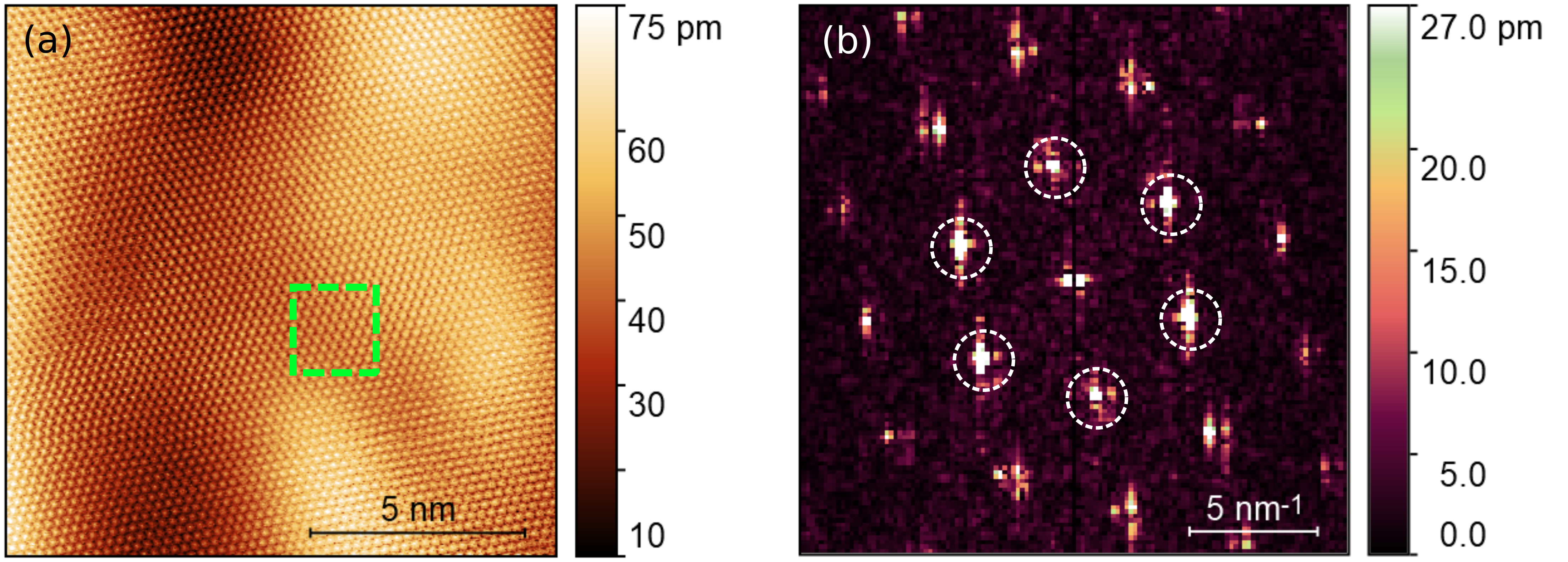}
    \caption{
      \textbf{Large area, atomic resolution topography image.}
      \textbf{(a)} Area measured in Fig. 5 of the main text shown by green, dashed rectangle.
      The only transformation of the raw image was a line-wise linear fit to the fast scan direction.
      The $\sqrt{3} \times \sqrt{3}$ pattern does not show up in this larger topographic image, which is unusual for defect induced scattering patterns.
      \textbf{(b)} Fourier transform of the image in (a).
      White, dashed circles show the Fourier components of the atomic lattice, similarly to Fig. 5d,e of the main text.
      The $\sqrt{3} \times \sqrt{3}$ modulation is not visible in the Fourier transform.
      Measurement parameters: 30 pA, 30 mV.
    }
  \label{fig:fig5extended}
  \end{center}
  \end{figure}

  Here we present a larger area topographic image, measured around the area of the images presented in Figure 5 of the main text.
  The image shows no signs of strong lattice defects, which is reinforced by the fact that there is no intervalley scattering peaks in the Fourier transform, see Fig.~\ref{fig:fig5extended}b.
  This is a strong signature that the $\sqrt{3} \times \sqrt{3}$ modulation observed in Fig. 5 of the main text does not originate from surface defects, such as vacancies.
  Nonetheless, subsurface defects can not be ruled out, as these would have a much weaker surface LDOS modulation~\cite{Dutreix2016-ye}.
  The defect origin of the LDOS modulation will be investigated in future devices, having a gate electrode.
  In such devices the sample charge density could be tuned to display various many-body ground states, allowing to switch the $\sqrt{3} \times \sqrt{3}$ pattern on and off by tuning the gate voltage.

\newpage

\section{Other models for surface state splitting}
\label{sec:other_models}

\subsection{Electric field induced surface state splitting}
\label{sec:theo_efield}
  
  In principle an electric field perpendicular to the graphene layers can open a gap in the surface state, as demonstrated in recent charge transport measurements \cite{Shi2020-bv}. 
  One important feature of this type of gap opening is that it does not necessarily occur at the Fermi level, such as in our case.
  Since the splitting we observe is always centered on the Fermi level, we can rule out this type of gap opening mechanism.

\begin{figure}
    \includegraphics[width = 0.9 \textwidth]{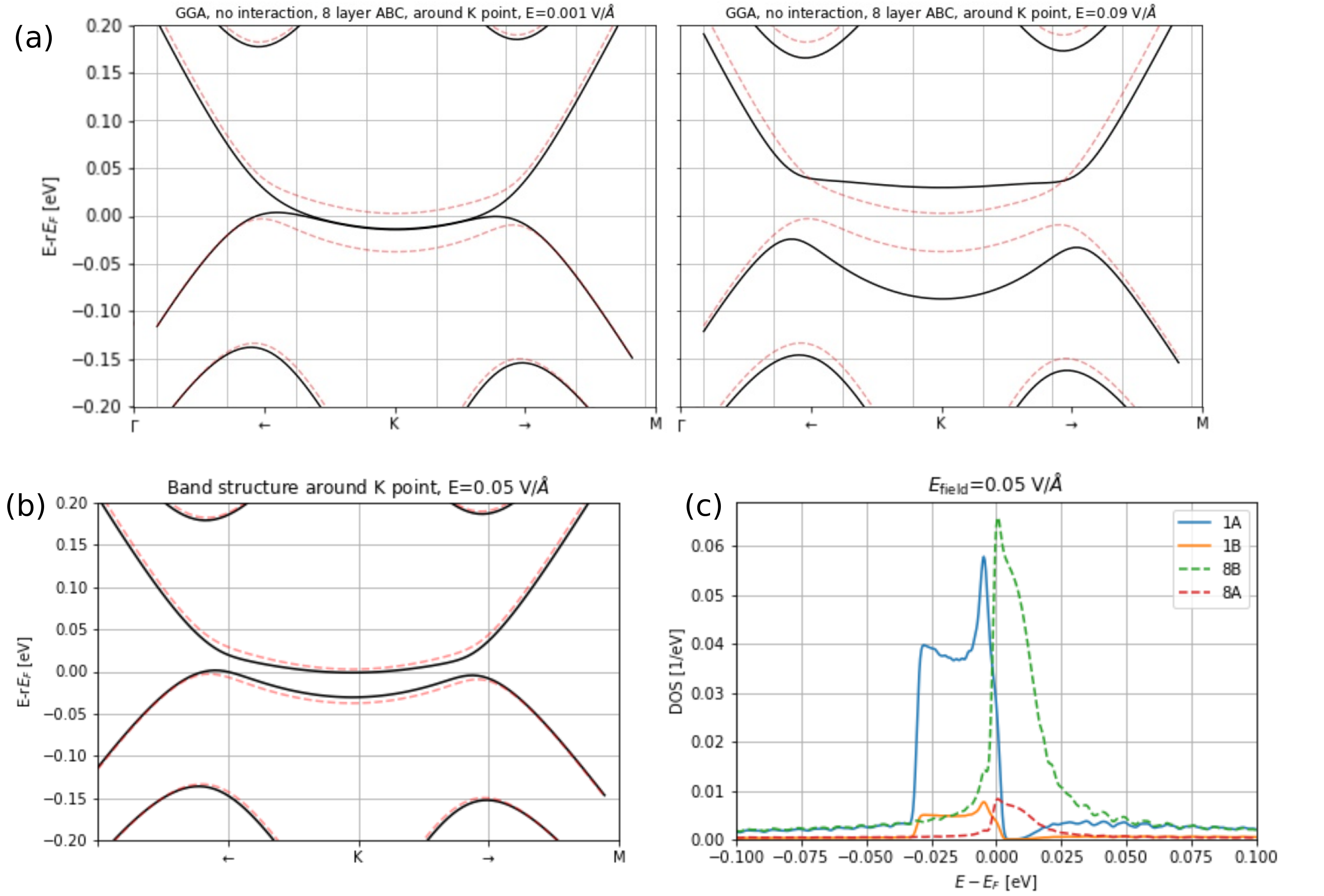}
    \caption{
      \textbf{Surface state splitting by a perpendicular electric field.}
      \textbf{(a)} Full black lines: band structure of an 8 layer RG, around the K point (SIESTA). Left: for no electric field. Right: Electric field of 0.09 V/\AA, perpendicular to to graphene layers.
      Dashed lines: splitting induced by including an effective Coulomb repulsion of 1.9 eV in the DFT calculation (see \ref{sec:calc_dft}).
      \textbf{(b)} The splitting due to interactions is close to the electric field induced one for an electric field of $\sim$0.05 V/\AA.
      \textbf{(c)} Total calculated DOS for an 8 layer thick slab, at a perpendicular electric field of 0.05 V/\AA. Curves 1A and 1B show the DOS on the top graphene layer, for sublattices A and B respectively. The curves 8A and 8B show the DOS on A and B sublattice of the bottom graphene layer. It is clear that the LDOS on the two sides of the split peak is localized to the top and bottom graphene layer of the RG crystal.
    }
  \label{fig:e-field}
  \end{figure}

  Investigating the problem from a more quantitative approach, we don't expect a significant electric field between the tip and the RG surface, since we are measuring in the vicinity of 0 bias voltage.
  Nevertheless we have investigated using DFT calculations the gap opening due to an out of plane electric field.
  For details see Fig. \ref{fig:e-field}, where we plot the band structure around the K point and the LDOS of the top graphene surface.
  An electric field of 0.09 V/\AA\ is necessary to open a surface state splitting of $\sim$30 meV, as observed in our experiment.
  This value is clearly not present in our experiment, since the maximum field reached in devices with top and bottom gates is 0.08 V/\AA\ \cite{Shi2020-bv}.

\subsection{Splitting of the surface state due to quantum confinement}
\label{sec:theo_confinement}

  Two dimensional materials on top of Si/SiO$_2$ experience a local doping modulation on the $\sim$20 nm length scale \cite{Zhang2009-gn,Jung2011-lk,Samaddar2016-zg}.
  This doping landscape can induce quantum confinement effects \cite{Zhao2015-dp,Lee2016-sj} that can modify the shape of the surface state and under certain circumstances induce splitting.
  In this section we present results of a detailed numerical study of the characteristic features of the peak observed in the local density of states (LDOS) corresponding to the flat band, obtained from a non-interacting tight binding model. We show that although under certain conditions a 
non-interacting model can also yield splitting of the peak, its distinctive features are markedly
different from the splitting caused by many body correlations discussed in the main text.


The results were obtained by performing kernel polinomial method \cite{KPM_paper} calculations of
large but finite systems. We considered, if otherwise not stated, in these calculations a cylindrical
shaped ABC stacked graphene sample with $R=1200$ \AA\ radius and 8 atomic layers thickness. We explored
the impact of an external potential profile $U(\mathbf{r})$ on the local density of states of the topmost layer.

\begin{figure}
    \centering
  \includegraphics[width = 0.7 \textwidth]{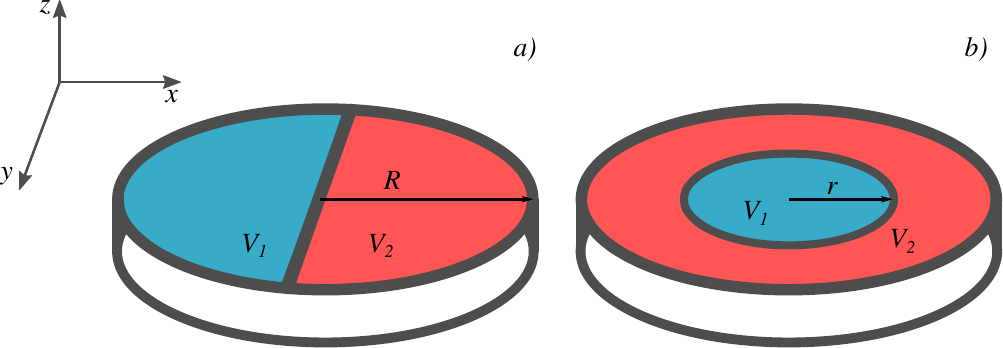}
  \caption{Considered geometries. a) a potential step, b) a circular potential well. 
  \label{fig:supmat_geometry}}
\end{figure}

We considered two potential configurations, depicted in Fig.~\ref{fig:supmat_geometry}, a) a symmetric
potential step perpendicular to the $x$ direction described by 
\begin{equation}
U(\mathbf{r})=U(x)= (V_2-V_1)\left(\arctan(x\beta)/\pi -1/2 \right )+V_1,
\end{equation}
where $x$ is measured from the center of the sample in the $x$ direction,
and b) a circular potential well specified by the expression
\begin{equation}
U(\mathbf{r})=U(r)=(V_2-V_1)\left[\arctan\left ( \left (r-R_{pot} \right )\beta\right)/\pi -1/2 \right ]+V_1
\end{equation}
where $r$ is the radial distance from the center of the sample.
In these expressions $V_1$, and $V_2$ control the height and symmetry of the potential profile, while 
the $\beta$ parameter specifies the slope. For both potential profiles we assumed that the potential is
independent of height, thus it is the same for all layers.

\begin{figure}
    \centering
  \includegraphics[width = 0.5 \textwidth]{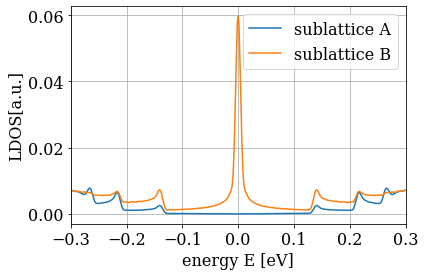}
  \caption{Local density of states of atoms on the A and B sublattice on the topmost layer in the center
  of the sample in the absence of an external potential. \label{fig:supmat_clean_AB}}
\end{figure}

In Fig.~\ref{fig:supmat_clean_AB} the sublattice polarization of the local density of states of the
central atoms in the topmost layer, in the absence of an external potential, is depicted. For $E=0$ eV
the local density of states has a prominent peak on sublattice A, while sublattice B is featureless for
small energies. For higher energies on both sublattices characteristic step like features are present
which are commensurate with the quantization due to the thickness of the sample.

\begin{figure}
    \centering
  \includegraphics[width = 0.8 \textwidth]{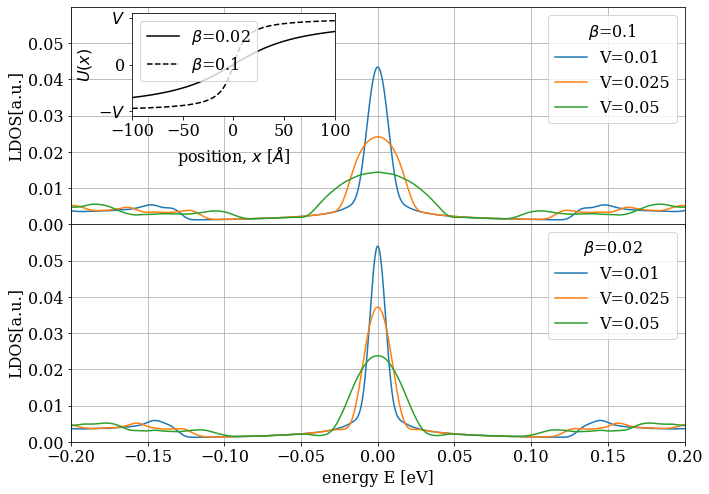}
  \caption{Local density of states in the presence of a potential step, for various step heights 
  $V$ and slopes $\beta$ at the central site. \label{fig:supmat_step_center}}

\end{figure}

Let us now consider step like potential configurations, such as those depicted in Fig.~\ref{fig:supmat_geometry} a). 
For these calculations we limit ourselves to symmetric potential configurations that is $V_1=-V_2=V$ since
the energy dependence of the local density of states for an asymmetric configuration can be deduced from
the symmetric one through a uniform shift in energy. Furthermore we shall also limit ourselves to the
discussion of the density of states on sublattice B. As it can be observed in  Fig.~\ref{fig:supmat_step_center}
at the charge neutrality point of the potential step the local density of states still displays a peak at zero energy
however, the height and width of the peak is influenced by the potential height and the slope of the potential step.
As expected the peak is higher and sharper for smaller potential steps and more gentle potential profiles.

\begin{figure}
    \centering
  \includegraphics[width = 0.8 \textwidth]{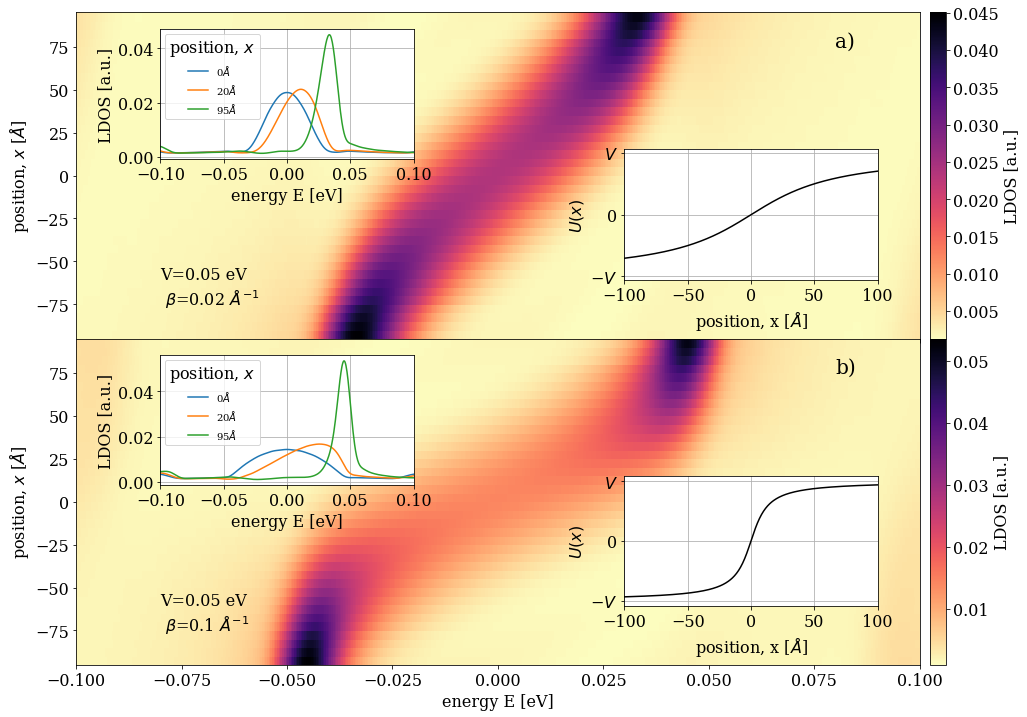}
  \caption{Local density of states in the presence of a potential step at different sites for a given
  potential step size of $V=0.05$ eV and for $\beta=0.02$ \AA$^{-1}$ in a) while $\beta=0.1$ \AA$^{-1}$ in b).
  \label{fig:supmat_step_sites}}

\end{figure}

Moving away from the interface the peak recovers its sharpness gradually but it also develops an asymmetric shape
as it can be observed in Fig.~\ref{fig:supmat_step_sites}. The asymmetry of the peak is enhanced for sharper
interfaces that is for higher $\beta$ values. 

\begin{figure}
    \centering
  \includegraphics[width = 0.8 \textwidth]{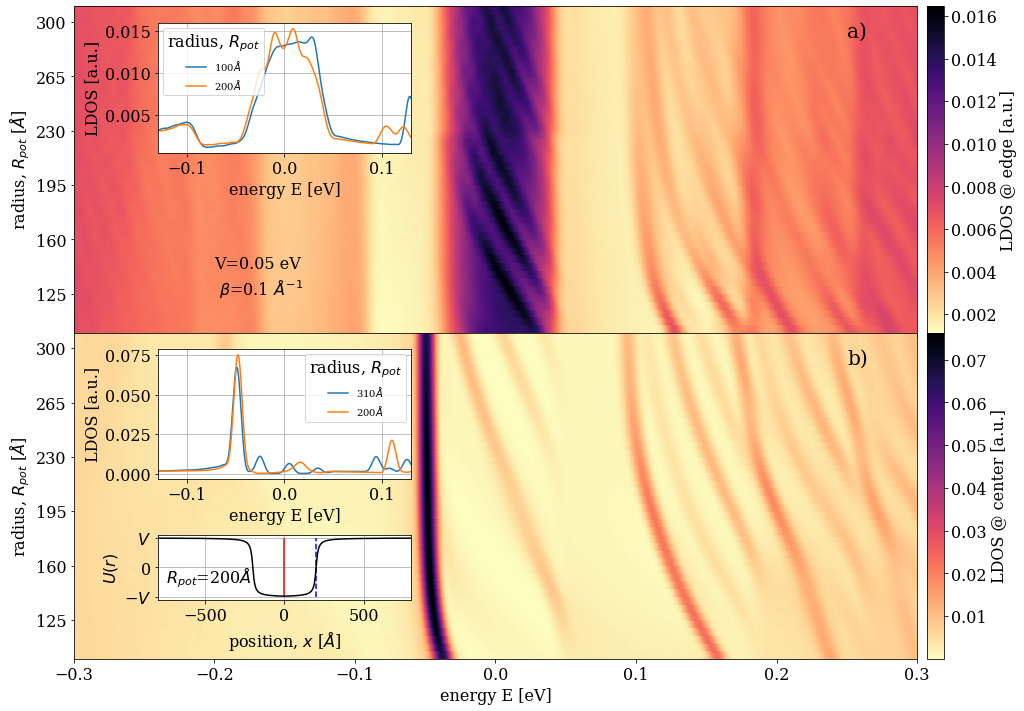}
  \caption{Local density of states on the surface of the sample in the presence of a circular potential well. 
  a) LDOS at the edge of the potential profile for various sizes of the potential well characterized by
  $R_{pot}$ b) LDOS in the center of the sample. Insets show sample crossections of the LDOS and the shape
  of the potential profile considered. Red solid line and blue dashed line denote the position of the 
  investigated site in subfigure b) and a) at $R_{pot}=200$ \AA\ respectively.
  \label{fig:supmat_circle}}
\end{figure}

Focusing now on the potential well geometry, as shown in Fig. \ref{fig:supmat_geometry} b), we again first consider
symmetric potential profiles that is $V_1=-V_2=V$ Fig.~\ref{fig:supmat_circle} 
depicts the local density of states for such a potential arrangement. 
In subfigure a) where the local density of states at the edge of the potential well is shown as a function of the
radius $R_{pot}$ a prominent peak can be observed close to zero energy. The top of the peak is modulated as $R_{pot}$
is increased giving rise to distinct double peak features for selected $R_{pot}$ values, as it can be observed in the
inset for $R_{pot}=200$ \AA. However note that the position of this double peak feature is not necessarily situated
symmetrically at zero energy, nor is a complete separation of the peak in to two distinct peaks observed, as it was
seen in the measurements detailed in the main text. Moreover for very small $R_{pot}$ the double peak structure is
not even visible, as it is seen in the inset for $R_{pot}=100$ \AA.
In b) the local density of states at the center of the potential well is shown as the function of $R_{pot}$. 
In this case the sharp peak corresponding to the flat band is shifted by the strength of the potential $V$ as expected
since around this position the system behaves as a uniformly doped sample. In both subfigures a distinct pattern tuned
by the radius $R_{pot}$ can be noted for positive energies as well.

\newpage

\begin{figure}
    \centering
  \includegraphics[width = 0.8 \textwidth]{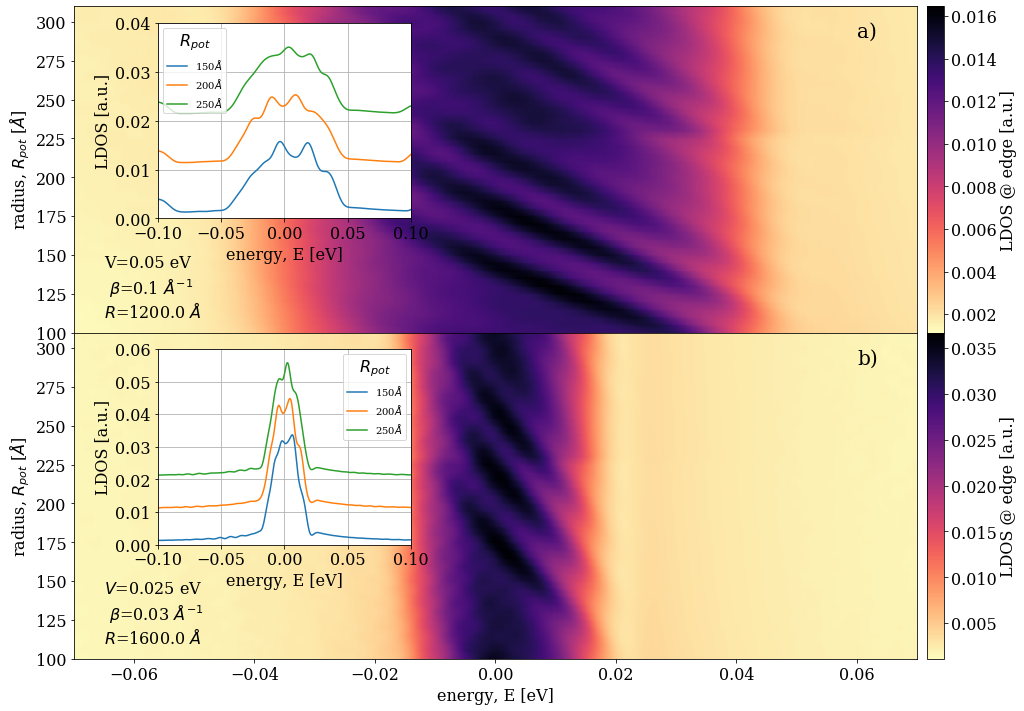}
  \caption{Local density of states on the surface of the sample in the presence of a circular potential well at the edge of the 
  well as a function  of the radius of the well $R_{pot}$ for two different potential configurations. In a) $V=0.05$ eV and 
  $\beta=0.1$ \AA$^{-1}$  the potential was steeper and reaches a higher value while in b) $V=0.025$ eV and $\beta=0.03$ \AA$^{-1}$
  that is the potential more gentle and of moderate height. 
  The sample size in a) was $R=1200$ \AA\ while in b) it was increased to $R=1600$ \AA\ in order to alleviate spurious 
  interference patterns due to finite size effects. Insets in both subfigures show representative crossections.
  \label{fig:supmat_circle_2}}
\end{figure}

\begin{figure}
    \centering
  \includegraphics[width = 0.8 \textwidth]{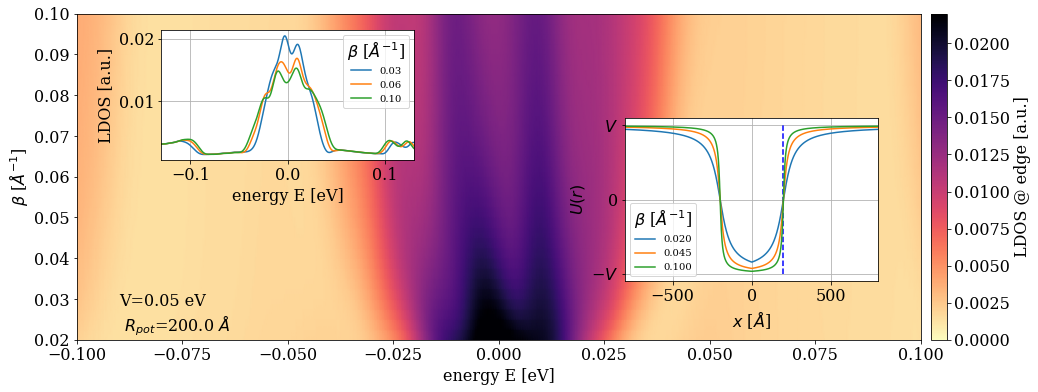}
  \caption{Local density of states on the surface of the sample in the presence of a circular potential well at the edge of the 
  well as a function of the slope parameter $\beta$. 
  \label{fig:supmat_circle_beta}}
\end{figure}

We explore how the height and steepness of the potential well impacts the fine structure of the central peak in 
Fig.~\ref{fig:supmat_circle_2}. As expected from the calculations for the simple potential step, the height of the
potential is directly linked to the width of the zero energy peak. The details of the modulation of the top of the peak 
due to changing $R_{pot}$ remains still distinguishable for smaller and more gentile potential steps. 
In Fig.~\ref{fig:supmat_circle_beta} the impact of the slope parameter $\beta$ on the details of the zero energy peak is explored. 
One can observe that although the overall shape of the peak is slightly modulated the positions of the two local maxima is largely
insensitive to the particular choice of $\beta$. 

\newpage

\begin{figure}
    \centering
  \includegraphics[width = 0.8 \textwidth]{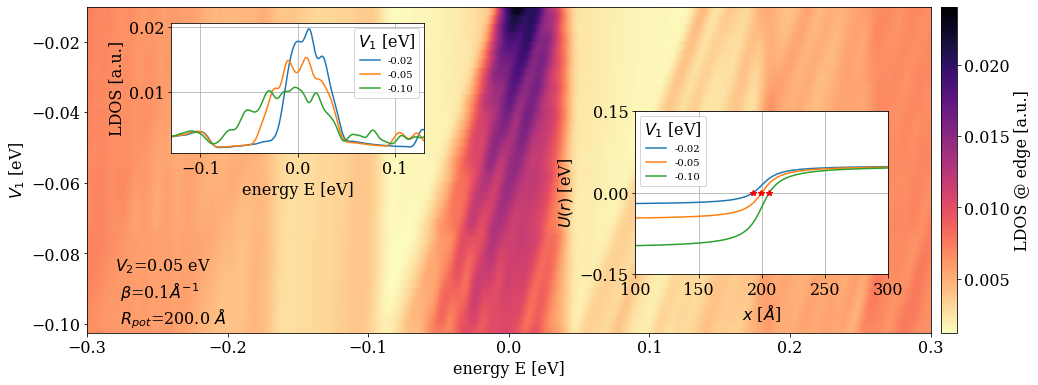}
  \caption{Local density of states on the surface of the sample in the presence of a circular potential well at the edge of the 
  well as a function  of the depth of the potential $V1$ inside the well for fixed potential values outside the well $V_2$. 
  The two insets show some characteristic crossections for the local density of states and the shape of the potential profile 
  considered for selected values of $V_1$. Red stars denote the positions of the sites where the local density of states was obtained. 
  \label{fig:supmat_circle_step}}
\end{figure}

Finally we explore how the details of the central peak are impacted for asymmetric potential wells where $V_1$ and $V_2$ are independent of each other.
These results are summarized in Fig.~\ref{fig:supmat_circle_step}. One can clearly observe that the position of the local maxima of the central peak 
linearly depend on $V_1$.  

\newpage

In this section we investigated whether the experimentally observed splitting of the central peak could be explained by interference
patterns of single particle excitations. We showed that under certain circumstances modulations of the central peak can be observed 
which could be interpreted as the splitting of the central peak. However detailed analysis shows that the characteristic features of
these modulations are markedly different than those observed in the measurements.

\newpage

\section{Atomic contrast of the surface state}
\label{sec:contrast}

\begin{figure}
  \begin{center}
    \includegraphics[width = 0.9 \textwidth]{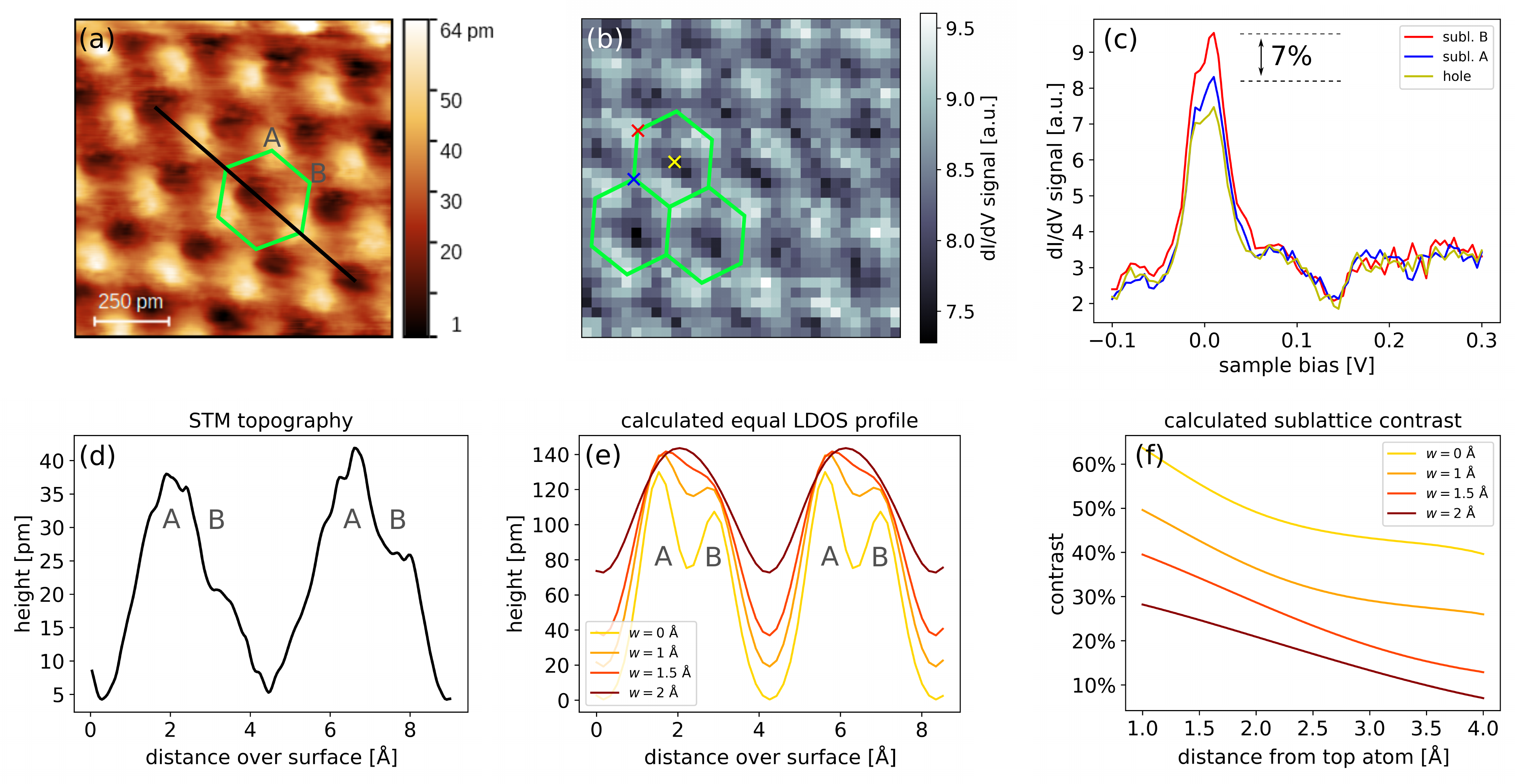}
    \caption{
      \textbf{Atomic contrast of the surface state.}
      \textbf{(a)} Topographic STM image. Stabilization parameters: 0.5 V, 300 pA. Black line shows the position of the height section shown in (d). Sublattices denoted "A" and "B".
      \textbf{(b)} A 32$\times$32 dI/dV map measured in the same area as (a). Stabilization parameters: 0.5 V, 300 pA. The color scale denotes the height of the surface state, resulting form a Gaussian fit to the peak.
      \textbf{(c)} Selected spectra in the positions marked by colored crosses in (b). The LDOS contrast of the surface state is calculated to be 7\%, using equation \ref{eq:contrast_exp}.
      \textbf{(d)} Height section of the STM topography, along the black line in (a).
      \textbf{(e)} Calculated (VASP) equal LDOS lines above the top graphene layer in a 15 layer RG crystal. The LDOS was integrated in an energy range between the Fermi level and 0.5 eV. Different colors mark the result of Gaussian smoothing, with a width $w$ between 0 and 2 \AA.
      \textbf{(f)} Calculated (VASP) surface state LDOS contrast ($C_{\mathrm{theo}}$) of the two sublattices, as a function of distance from the surface of RG.
    }
  \end{center}
  \label{fig:contrast}
  \end{figure}

  The surface state of RG is localized to the unpaired sublattice of the top graphene layer, shown in blue in Fig. 1a of the main text.
  Therefore, there should be a difference in the measured local density of states (LDOS) on the two sublattices.
  In Fig. \ref{fig:contrast}a, we present a typical atomic resolution STM image of the sample surface, showing the two sublattices of the top graphene layer, denoted by "A" and "B".
  Since the surface state is near the Fermi level, ideally at 0 V bias voltage, it is difficult to measure topographic STM images at voltages near 0 V.
  Therefore we have measured a 32$\times$32 $dI/dV$ map in the area shown in Fig. \ref{fig:contrast}a, using a reduced voltage range in order to increase the measurement speed and minimize the effect of thermal drift.

  In Fig. \ref{fig:contrast}b we plot the $dI/dV$ signal intensity of the surface state, resulting from a Gaussian fit to the peak.
  By examining this map, it is clear that the intensity of the surface state varies on an atomic scale, with the peak intensity being lower on the A sublattice (see Fig. \ref{fig:contrast}c).
  The tip-sample distance in this measurement was stabilized at 0.5 V.
  By using this measurement protocol, we make sure that the tip-sample distance is determined by the LDOS of the bulk states up to 0.5 eV and we probe the surface state LDOS variation at this separation.
  By examining the individual spectra it is clear that the bulk states in the 0.2 to 0.3 V range have the same $dI/dV$ signal, whereas the only difference between the spectra is in the intensity of the surface state peak.
  The experimental sublattice contrast ($C_{\mathrm{exp}}$) is calculated using the following relationship:
  
  \begin{equation}
    C_{\mathrm{exp}} = 100 \frac{(dI/dV)_{\mathrm{A}} - (dI/dV)_{\mathrm{B}}}{(dI/dV)_{\mathrm{A}} + (dI/dV)_{\mathrm{B}}},
    \label{eq:contrast_exp}
  \end{equation}

  where ($dI/dV_\mathrm{{A}}$) and ($dI/dV_\mathrm{{B}}$) denote the measured tunneling conductance maximum of the surface state, measured at a tip - surface distance set by the STM stabilization parameters of 300 pA and 0.5 V.
  Using equation \ref{eq:contrast_exp}, we find a sublattice contrast of 7\%.
  In similar measurements, the contrast was found to be between 5 and 10\%.
  On first glance this value seems to be low, but can be explained by considering the atomic arrangement in RG's top two graphene layers and by the fact that in the experiment the STM tip never tunnels only into one sublattice.

  Considering the first effect, we should notice that in the second atomic layer of RG, the sublattice with significant surface state LDOS resides under the top atom with near zero LDOS.
  Therefore, there is a non zero overlap of the surface state orbital in the second layer, with the STM tip orbital.
  This effect reduces the measured contrast below 100\%.
  Using DFT calculations (VASP), which include both contrast reducing effects mentioned above, we are able to qualitatively reproduce the measured contrast.
  Firstly we calculated the LDOS ($\rho$) in the energy range between the Fermi level and 0.5 eV for a 15 layer thick RG crystal.
  A contour of equal LDOS above the top graphene layer, along the armchair direction (similar to the section in Fig. \ref{fig:contrast}a) is shown by the yellow curve in Fig. \ref{fig:contrast}e.
  This plot best represents the topographic STM image measured at a bias of 0.5 V.
  To estimate the broadening effect of the tip, we smoothed the iso-LDOS profile using a Gaussian function, with a full width at half maximum ($w$) between 1 and 2 \AA.
  By comparing Fig. \ref{fig:contrast}d and \ref{fig:contrast}e we conclude that the broadening of $\sim$1.5 \AA\ is appropriate for this particular tip.
  Next we define the theoretical contrast ($C_{\mathrm{theo}}$) in a similar fashion to the experimental one:
  \begin{equation}
    C_{\mathrm{theo}} = 100 \frac{\rho_{\mathrm{A}} - \rho_{\mathrm{B}}}{\rho_{\mathrm{A}} + \rho_{\mathrm{B}}},
    \label{eq:contrast_theo}
  \end{equation}
  with the LDOS values $\rho_{\mathrm{A}}$ and $\rho_{\mathrm{B}}$ being the LDOS of the surface state.
  The values for $\rho_{\mathrm{A}}$ and $\rho_{\mathrm{B}}$ are calculated at a distance above the graphene A and B sublattices, determined by the equi-LDOS profile shown in Fig. \ref{fig:contrast}e.
  This takes into account the different tip - sample distance of the STM tip on each sublattice.
  We plot $C_{\mathrm{theo}}$ as a function of distance perpendicular to the top graphene layer in Fig. \ref{fig:contrast}f.
  We can see a decreasing trend for the sublattice contrast as we move away from the RG surface, with the contrast being of the order of 10\% for realistic tip - sample distances of $\sim$5 \AA.

\end{document}